\documentclass{article}

\usepackage{arxiv}
\usepackage[utf8]{inputenc} 
\usepackage[T1]{fontenc}    
\usepackage{hyperref}       
\usepackage{url}            
\usepackage{booktabs}       
\usepackage{amsfonts}       
\usepackage{nicefrac}       
\usepackage{microtype}      
\usepackage{lipsum}
\usepackage{geometry}
\usepackage{amsmath}
\usepackage{amssymb} 
\usepackage{graphicx}
\usepackage{mathrsfs}
\usepackage{cite}
\usepackage{color}
\usepackage{subfigure}
\usepackage{soul}
\usepackage[title]{appendix} 

\definecolor{nec}{RGB}{255,0,0}
\definecolor{sug}{RGB}{0,153,76}

\title{Bilayer graphene in magnetic fields generated by supersymmetry}
\author{
David J. Fern{\'a}ndez C., Juan D. Garc{\'i}a M. and Daniel O-Campa\\
Physics Department, Cinvestav\\
P.O.B. 14-740, 07000 Mexico City, Mexico\\
e-mail: david@fis.cinvestav.mx, dgarcia@fis.cinvestav.mx, dortiz@fis.cinvestav.mx
}

\begin{document}

\maketitle

\begin{abstract}
The effective Hamiltonian for electrons in bilayer graphene with applied magnetic fields is solved through second-order supersymmetric quantum mechanics. This method transforms the corresponding eigenvalue problem into two intertwined one dimensional stationary Schr{\"o}dinger equations whose potentials are determined by choosing at most two seed solutions. 
In this paper new kinds of magnetic fields associated to non-shape-invariant SUSY partner potentials are generated. Analytic solutions for these magnetic fields are found, the associated spectrum is analyzed, and the probability and current densities are explored.
\end{abstract}


\section{Introduction}
Bilayer graphene is a material composed of two stacked graphene layers, which can exist in nature either in the AB (so-called Bernal stacking) or in the AA forms, the first being the most common one \cite{Katsnelson2007}. Bilayer graphene is currently studied in many research areas, since its electronic properties turn out to be similar to the monolayer ones. However, for bilayer graphene at low energies the integer quantum Hall effect indicates the presence of massive chiral quasiparticles with a parabolic dispersion relation instead of the linear relation appearing for monolayer graphene \cite{McCann2013}. 

If dimer processes and low energies are considered, the effective Hamiltonian for bilayer graphene is not given by the Dirac-Weyl equation, as in monolayer graphene, but rather by an equation of second-order in the momentum. It is worth to recall that exact analytical solutions for monolayer graphene in external magnetic fields can be obtained through first-order supersymmectric quantum mechanics (SUSY QM) \cite{Kuru2009,Milpas2011,Midya2014,Erik2017,Concha2018,Roy2018,Erik2019,Celeita2020}. Furthermore, a recent study has shown that the second-order SUSY QM is the appropriate tool to solve the electron motion in bilayer graphene with magnetic fields which are related to shape-invariant SUSY partner potentials \cite{fgo20} (see also \cite{fm20}). The key point of this technique is to connect the effective Hamiltonian of bilayer graphene with two Schr{\"o}dinger equations intertwined by a second-order differential operator \cite{Andrianov1993,Andrianov1995,Samsonov1999,Salinas2003,Salinas2005,Nicolas2005,Fernandez2010,Fernandez2019,Barnana2020}.  We must stress that the method introduced in \cite{fgo20} was used to solve the effective Hamiltonian for bilayer graphene in the same sequence of magnetic fields addressed in \cite{Kuru2009}. In this article we are going to generalize the method, by showing that the second-order SUSY QM can be used as well when non shape-invariant SUSY partner potentials are involved.

This paper has been organized as follows: in section 2 we will discuss the effective Hamiltonian for bilayer graphene and the way to implement the second-order SUSY QM for solving the associated eigenvalue problem. In sections 3 and 4 new solvable cases generated by two variants of the method will be addressed, and some physical quantities will be analyzed. In section 5 we will present our conclusions.


\section{Effective Hamiltonian and second-order SUSY QM} \label{section2}
In the tight-binding model \cite{Mermin1976,Saito1998,Raza2012} the effective Hamiltonian for bilayer graphene turns out to be
\begin{equation}
H=\frac{1}{2m^{*}}
\begin{pmatrix} 
0 &\left(p_x-ip_y\right)^{2}\\
\left(p_x+ip_y\right)^{2} & 0
\end{pmatrix},
\label{2.1}
\end{equation}
which is neither a Dirac-like (relativistic case) nor a Schrödinger-like (non-relativistic case) effective Hamiltonian\cite{Katsnelson2007}.

If we apply magnetic fields which are orthogonal to the graphene surface ($x-y$ plane) and change only along a fixed given direction ($x$), the vector potential in the Landau gauge can be chosen as $\vec{A}=\mathcal{A}(x)\hat{e}_y$ and then $\vec{B}=\mathcal{B}(x)\hat{e}_z$, $\mathcal{B}(x)=\mathcal{A}'(x)$. Now, the minimal coupling rule transforms $p_i$ into $p_i+\frac{e}{c}A_i$, thus the eigenvalue equation for the Hamiltonian \eqref{2.1} in the applied magnetic field becomes
\begin{align}
H\Psi\left(x,y\right)&=\frac{1}{2m^{*}}
\begin{pmatrix} 
0 & \Pi^2\\
\left(\Pi^{\dagger}\right)^{2} & 0
\end{pmatrix}
\Psi\left(x,y\right) =E\Psi\left(x,y\right),
\label{2.2}
\end{align} 
where $\Pi=p_x-ip_y-i\frac{e}{c}\mathcal{A}(x)$. The invariance of this equation under translations along $y$-direction suggests the following expression for $\Psi\left(x,y\right)$: 
\begin{equation}
\Psi\left(x,y\right)=\frac{e^{iky}}{\sqrt{2}}
\begin{pmatrix} 
\psi^{\left(2\right)}\left(x\right)\\ 
\psi^{\left(0\right)}\left(x\right)
\end{pmatrix}, 
\label{2.3}
\end{equation}
with $k$ being the wave-number in $y$-direction. By plugging equation \eqref{2.3} into equation \eqref{2.2}, after some algebra it is obtained a coupled system of equations of the form
\begin{align}
L^{-}_2\psi^{\left(0\right)}\left(x\right)&=
-\widetilde{E}\psi^{\left(2\right)}\left(x\right),\label{2.4} \\
L^{+}_2\psi^{\left(2\right)}\left(x\right)&=
-\widetilde{E}\psi^{\left(0\right)}\left(x\right),\label{2.5}
\end{align}
where $L^{-}_2$ and $L^{+}_2=\left(L^{-}_2\right)^{\dagger}$ are given by
\begin{eqnarray}
L^{-}_2 & = & \frac{d^2}{dx^2}+\eta\left(x\right)\frac{d}{dx}+\gamma\left(x\right), \label{2.6} \\
L^{+}_2 & = & \frac{d^2}{dx^2}-\eta\left(x\right)\frac{d}{dx}+\gamma\left(x\right)-\eta'\left(x\right),
\nonumber 
\end{eqnarray} 
with $\eta$ and $\widetilde{E}$ being expressed as
\begin{equation}
\widetilde{E}=\frac{2m^{*}E}{\hbar^{2}},\quad\eta\left(x\right)=2\left(k+\frac{e}{c\hbar}\mathcal{A}(x)\right).
\label{2.7}
\end{equation}

Note that $\eta(x)$ is related linearly with the vector potential amplitude $\mathcal{A}(x)$, hence with the magnetic field as 
\begin{equation} \label{2.8}
\mathcal{B}(x)=\frac{c\hbar}{2e}\eta'(x). 
\end{equation}
The function $\gamma$ is related to $\eta$, thus with $\mathcal{A}$, through $\gamma = \eta'/2 + \eta^2/4$, so that $L^{-}_2 = (\frac{d}{dx} + \frac{\eta}2)^2$. However, in order to widen our options, including the possibility that $L^{-}_2$ would be expressed as the product of two in general different first-order differential operators, from now on two assumptions will be made: (i) $\eta$ and $\mathcal{A}$ will be neccessarily related by equation~(\ref{2.7}); (ii) $\eta$ and $\gamma$ will be as well related, but in such a way that $L^{\pm}_2$ will intertwine two Hermitian Schr\"odinger Hamiltonians, which implies that equation~(\ref{2.13}) below will be always fulfilled. These mathematical assumptions have to do physically with the inclusion of an extra term in the Hamiltonian of equation~(\ref{2.2}) in the following way: 
\begin{equation} \label{extra2}
H =\frac{1}{2m^{*}}
\left(\begin{array}{cc}
0 & \Pi^2\\
\left(\Pi^{\dagger}\right)^{2} & 0
\end{array}\right)
-\frac{\hbar^2}{2m^*} f(x) \, \sigma_{x},
\end{equation}
where
$$
f(x) = \frac{\eta'(x)^{2}}{4\eta^{2}(x)} - \frac{\eta''(x)}{2\eta(x)} - \frac{(\epsilon_{1} - \epsilon_{2})^{2}}{4\eta^{2}(x)}.
$$
This extra term could be associated to trigonal warping effects, spatially varying external potentials, among others \cite{Katsnelson2007,Wu2012}. We must also especially mention the phenomenon caused by strained deformation in graphene, whose mathematical description could be very close to the one given by equations~(\ref{2.4}), (\ref{2.5}) and (\ref{extra2}), with the assumptions posed in (i) and (ii).

Now, it is straightforward to decouple the system of equations (\ref{2.4}) and (\ref{2.5}) to get
\begin{align}
L_2^{+}L_2^{-}\psi^{(0)}(x)&=\widetilde{E}^{2}\psi^{(0)}(x), \nonumber\\
L_2^{-}L_2^{+}\psi^{(2)}(x)&=\widetilde{E}^{2}\psi^{(2)}(x). \label{2.9}
\end{align}
Since $L_2^{+}L_2^{-}$  and $L_2^{-}L_2^{+}$ are hermitian fourth-order differential operators, it seems natural to use the second-order supersymmetric quantum mechanics (SUSY QM) for dealing with the problem of bilayer graphene in external magnetic fields as follows.

Let us suppose that $\psi_n^{(l)}(x)$ with $\ l=0,2$ are the eigenfunctions of two Sch\"odinger Hamiltonians
\begin{equation}
H_l=-\frac{d^2}{dx^2}+V_l(x),
\label{2.10}
\end{equation}
associated to the eigenvalues $\mathcal{E}^{(l)}_n$. It is assumed that both Hamiltonians $H_l$ are intertwined, namely,
\begin{equation}
H_2L_2^{-}=L_2^{-}H_0.
\label{2.11}
\end{equation}  
The two $V_l$ are called SUSY partner potentials and $L_2^{-}$ is the intertwining operator appearing in equation~\eqref{2.6}. If equations~(\ref{2.6}) and (\ref{2.10}) are substituted in equation~(\ref{2.11}), after some algebra we get that 
\begin{align}
&V_2(x)=V_0(x)+2\eta'(x), \label{2.12}
\\
&\gamma(x)= \frac{\eta'(x)}{2} + \frac{\eta^2(x)}{4} - \frac{\eta''(x)}{2\eta(x)} + \left(\frac{\eta'(x)}{2\eta(x)}\right)^{2} - \left(\frac{\epsilon_{1} - \epsilon_{2}}{2\eta(x)}\right)^{2}, \label{2.13}
\\
&V_0(x)=\frac{\eta''(x)}{2\eta(x)}-\bigg(\frac{\eta'(x)}{2\eta(x)}\bigg)^2-\eta'(x)+\frac{\eta^2(x)}{4}+\bigg(\frac{\epsilon_1+\epsilon_2}{2}\bigg)+\bigg(\frac{\epsilon_1-\epsilon_2}{2\eta(x)}\bigg)^2, \label{2.14}
\end{align}   
where $\epsilon_1$ and $\epsilon_2$ are called factorization energies, which are associated to the two seed solutions $u_m^{(0)}	(x),\ m=1,2$ that belong to the kernel of $L_2^{-}$ and simultaneously are formal eigenfunctions of $H_0$, $L_2^{-}u_m^{(0)}=0$, $H_0u_m^{(0)}=\epsilon_m u_m^{(0)}$. In general, $\epsilon_1$ and $\epsilon_2$ can be complex, but we will restrict ourselves to cases where they will be real. Moreover, in this paper they will be chosen as two eigenvalues $\mathcal{E}^{(0)}_n$ associated to the corresponding eigenfunctions $\psi_n^{(0)}(x)$ of $H_0$. Whenever that $\eta$ is given, the eigenfunctions $\psi_n^{(2)}$ of $H_2$ can be calculated as follows,
\begin{equation}
\psi_n^{(2)}(x)=\frac{L_2^{-}\psi_n^{(0)}(x)}{\sqrt{(\mathcal{E}_n^{(0)}-\epsilon_1)(\mathcal{E}_n^{(0)}-\epsilon_{2})}},
\label{2.15}
\end{equation} 
since equations (\ref{2.11}~-~\ref{2.14}) guarantee that the products of $L_{2}^{\pm}$ become a second-degree polynomial in $H_{l}$, thus when applied to the eigenfunctions $\psi_{n}^{(l)}(x)$ it turns out that \cite{Nicolas2005}  
\begin{eqnarray} \label{extra1}
\nonumber L_{2}^{+}L_{2}^{-}\psi_{n}^{(0)}(x) = \left(H_{0} - \epsilon_{1}\right)\left(H_{0} - \epsilon_{2}\right)\psi_{n}^{(0)}(x) = \left(\mathcal{E}_{n}^{(0)} - \epsilon_{1}\right) \left(\mathcal{E}_{n}^{(0)} - \epsilon_{2}\right)\psi_{n}^{(0)}(x), \\
L_{2}^{-}L_{2}^{+}\psi_{n}^{(2)}(x) = \left(H_{2} - \epsilon_{1}\right)\left(H_{2} - \epsilon_{2}\right)\psi_{n}^{(2)}(x) = \left(\mathcal{E}_{n}^{(0)} - \epsilon_{1}\right) \left(\mathcal{E}_{n}^{(0)} - \epsilon_{2}\right)\psi_{n}^{(2)}(x).
\end{eqnarray}

According to the previous second-order SUSY treatment, the function $\eta(x)$ is calculated from the two seed solutions of $H_{0}$, so the choice of  factorization energies is relevant. In the next section we will focuss on two cases of the second-order SUSY QM, for specific choices of factorization energies such that the SUSY partner potentials $V_0(x)$ and $V_2(x)$ are no longer shape-invariant. Moreover, the eigenfunctions and eigenvalues of the original problem are going to be found by comparing equation \eqref{2.9} with \eqref{extra1}; we shall give their explicit forms in the corresponding section for the two cases mentioned before. We must also note that although $V_0$ and $V_2$ are related to the external magnetic fields, they have no physical meaning so they are only auxiliary tools to solve the eigenvalue problem we are interested in.

Substituting now equation \eqref{2.8} in \eqref{2.12} it is straightforward to see that
\begin{equation} \label{2.16}
V_2(x)=V_{0}(x)+4\frac{e}{c\hbar}\mathcal{B}(x).
\end{equation}
Furthermore, the probability and current densities to be studied are given by
\begin{equation}
\rho(x,y)=\Psi^{\dagger}\Psi,
\label{2.17}
\end{equation}
and
\begin{eqnarray}
& J_{\ell}(x,y)=\frac{\hbar}{m^{*}}{\rm Im}\left(\Psi^{\dagger} \, j_{\ell} \, \Psi\right) + \frac{e}{m^{*}c}\mathcal{A}(x)\Psi^{\dagger}\varsigma_{\ell}\Psi, \quad \ell=x,y, \label{2.18} \\
& j_{x}=\sigma_x\partial_x+\sigma_y\partial_y,\quad
j_{y}=\sigma_y\partial_x-\sigma_x\partial_y,\quad \varsigma_{\ell} = \varepsilon_{\ell m}\sigma_{m},\label{2.19}
\end{eqnarray}
regardless of the factorization energies choice (see \ref{apendice2}) with $\varepsilon_{\ell m}$ being the Levi-Civita symbol in two dimensions. Note that the current density (\ref{2.18}) is not the same as the expression for the free case (without magnetic field) shown in \cite{Ferreira2011}. In fact, we must say that in \cite{fgo20} the second term of equation~(\ref{2.18}) was missing in the corresponding expression, a mistake that we are correcting in this paper.

\section{Factorization energies as two consecutive energy levels}
Given the potential $V_0$ and its corresponding eigenfunctions $\psi_n^{(0)}$ and eigenvalues $\mathcal{E}^{(0)}_n$, we choose now the factorization energies as two consecutive eigenvalues of $H_0$, i.e., $\epsilon_2=\mathcal{E}^{(0)}_j$ and $\epsilon_1=\mathcal{E}^{(0)}_{j+1}$, where $j\geq 1$. This choice supplies the following $\eta$-function
\begin{equation}
\eta (x)=-\frac{W'(\psi_{j}^{(0)},\psi_{j+1}^{(0)})}{W(\psi_{j}^{(0)},\psi_{j+1}^{(0)})},
\label{3.1}
\end{equation}
where $W(f,g)$ denotes the Wronskian of $f$ and $g$. 
Even though $V_0$ is the same as in the shape-invariant case \cite{fgo20}, which is recovered here for $j=0$, now the magnetic field calculated through equation~(\ref{2.8}) in general will change, becoming the sum of the shape-invariant part plus a rational modification. Similarly, the potential $V_2$ turns out to be the addition of the shape-invariant SUSY partner potential plus a term that involves the previously mentioned ratio.

The corresponding eigenvalues for the electrons in bilayer graphene turn out to be
\begin{equation}
E_{n}=\frac{\hbar^{2}}{2m^{*}}\sqrt{\Delta_{n,j}\Delta_{n,j+1}},\qquad
n=0,1,\dots
\label{3.2}
\end{equation}
where $\Delta_{n,j}\equiv\mathcal{E}^{(0)}_{n}-\mathcal{E}^{(0)}_{j}$, while the corresponding eigenfunctions are
\begin{equation}
\Psi_{n}(x,y) =
\begin{cases}
e^{iky}
\begin{pmatrix}
0\\
\psi_n^{(0)}(x)
\end{pmatrix}
& \text{for}\ n= j, j+1, \\
\frac{e^{iky}}{\sqrt{2}}
\begin{pmatrix}
\psi_{n}^{(2)}(x)\\
\psi_n^{(0)}(x)
\end{pmatrix}
& \text{for}\ n\neq j, j+1.
\end{cases}
\label{3.3}
\end{equation}

Let us note that the index $n$ supplies the standard ordering for the eigenvalues of the auxiliary problem, namely, $\mathcal{E}^{(0)}_{n}<\mathcal{E}^{(0)}_{n+1} \ \forall \ n=0,1,\dots$. Nevertheless, this does not happen in general for the electron energies $E_n$ in bilayer graphene, i.e., the set $\{E_n|n=0,1...\}$ is not ordered in the standard way. It would be necessary to define a new index for getting the standard ordering of $E_n$, but that will depend on the election of $j$ and the parameters involved in the potential $V_0$.

In the following examples we will study how the magnetic field is deformed when we choose two consecutive eigenenergies of $H_0$, with $V_0$ taken as in the shape-invariant case \cite{fgo20}. It means that, departing from $V_0$ we will determine both, the new potential $V_2$ and the magnetic field leading to these SUSY partner potentials. In order to simplify the discussion, from now on we will fix $j=1$, i.e., the first and second excited states will be taken as seed solutions for implementing the second-order SUSY technique.

\subsection{Shifted harmonic oscillator potential}
Let us take first the  potential $V_0$ as
\begin{equation}
V_{0}(x)=\frac{\omega^2}{4}\bigg(x+\frac{2\kappa}{\omega}\bigg)^2-\frac{\omega}{2},
\label{3.1.1}
\end{equation}
which is a shifted harmonic oscillator potential, where $\omega$ and $\kappa$ are real parameters such that $\kappa$ will be a function of the wavenumber $k$. Its eigenfunctions $\psi_n^{(0)}$ and eigenvalues $\mathcal{E}_n^{(0)}$ are given by
\begin{eqnarray}
& \psi_{n}^{(0)}(\zeta) = c_n e^{-\frac{1}{2}\zeta^2}H_{n}(\zeta),
\label{3.1.3} \\
& \mathcal{E}_{n}^{(0)} = n\omega, \quad n=0,1,2,\dots,
\label{3.1.2}
\end{eqnarray}
where $c_n$ is a normalization factor and $\zeta=\sqrt{\omega/2}(x+2\kappa/\omega)$. 

If the seed solutions and factorization energies are taken as mentioned before, 
$\epsilon_1=\mathcal{E}^{(0)}_{2}$ and $\epsilon_2=\mathcal{E}^{(0)}_1$, when substituting them in equation~(\ref{3.1}) we will get the corresponding function $\eta(x)$ which in turn will allow us to determine the SUSY partner potential $V_2$ and the corresponding magnetic field through equations~(\ref{2.16}) and (\ref{2.8}), namely,
\begin{eqnarray} 
&& V_2(x)=V_{SI}(\zeta)+4\omega\frac{2\zeta^2-1}{(2\zeta^2+1)^2}, \label{3.1.4}\\ 
&& \mathcal{B}(x)=\mathcal{B}_{SI}(\zeta)\left[1+\frac{4\zeta^2-2}{(2\zeta^2+1)^2}\right].
\label{3.1.5}
\end{eqnarray}
Note that we have obtained now a non-constant magnetic field $\mathcal{B}(x)$, in contrast with the shape-invariant case for which $\mathcal{B}_{SI}(\zeta)=\frac{c\hbar\omega}{2e}$ and $V_{SI}(\zeta)=(\omega/2)\zeta^2+(3/2)\omega$ \cite{fgo20}. Moreover, by taking into account equations (\ref{2.7}) and (\ref{3.1.5}) we get as well that $\kappa=k$ (see Appendix \ref{apendice}), where $k$ is the wave number in $y$-direction . The corresponding electron energies in bilayer graphene under the magnetic field of equation~(\ref{3.1.5}) are
\begin{equation}
E_{n}=\frac{\hbar ^2\omega}{2 m^*}\sqrt{(n-1) (n-2)}.
\label{3.1.6}
\end{equation}
Once again, it is important to observe that the energies $E_n$ do not have the same standard ordering as the auxiliary energies $\mathcal{E}_n^{(0)}$ (compare equations~(\ref{3.1.2}) and (\ref{3.1.6})). However, for this particular case it is easy to determine the standard ordering of these eigenvalues. In fact, for an arbitrary $j$ the electron energies are proportional to the square root of $\left(n-j\right)\left(n-j-1\right)$. Thus, the pairs $\{\left(\Psi_j,\Psi_{j+1}\right),\left(\Psi_{j-1},\Psi_{j+2}\right),\dots,\left(\Psi_{0},\Psi_{2j+1}\right)\}$ have ordered growing energies which are different and doubly degenerate, with the first pair corresponding to the ground state energy, the second pair to the first excited state energy and so on until the $j$-th excited state energy. Finally, the states $\{\Psi_{2j+2},\Psi_{2j+3},\dots\}$ are associated to eigenvalues $\{E_{2j+2},E_{2j+3},...\}$ which are non-degenerate, with $E_{2j+2}$ being the energy for the $(j+1)$-th excited state, and so on. For our example we have fixed $j=1$, so we have $2$ twofold degenerate levels corresponding to the ground state energy ($n=1,2$) and the first excited state energy ($n=0,3$), in contrast with the shape-invariant case where the only twofold degenerate eigenvalue is the zero energy ground state.

\begin{figure}[ht] 
\begin{center}
\subfigure[]{\includegraphics[width=6.97cm, height=4.675cm]{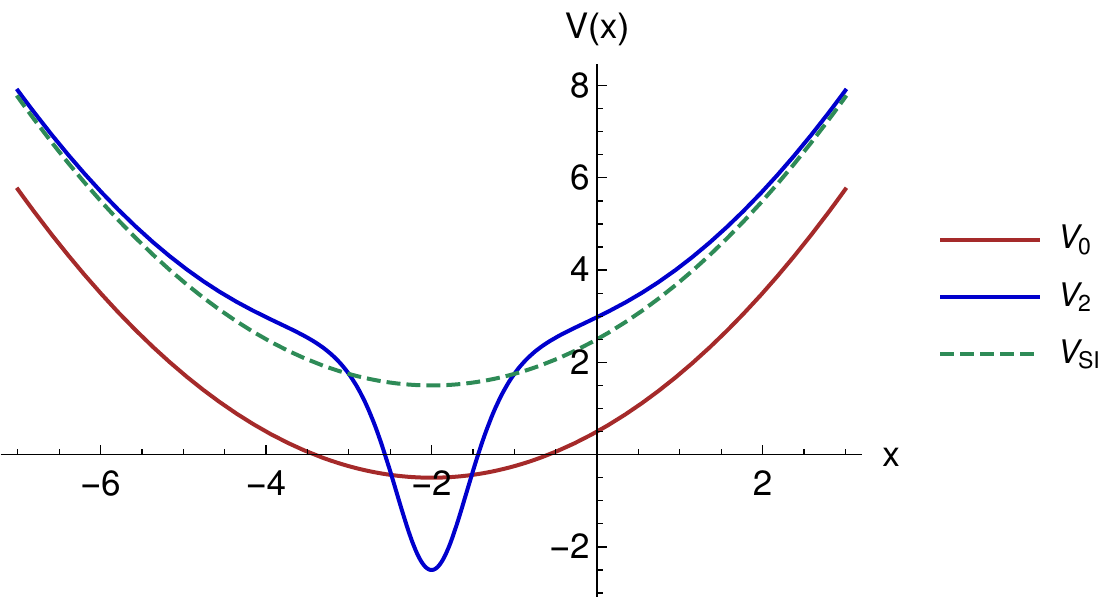}}
\subfigure[]{\includegraphics[width=6.97cm, height=4.675cm]{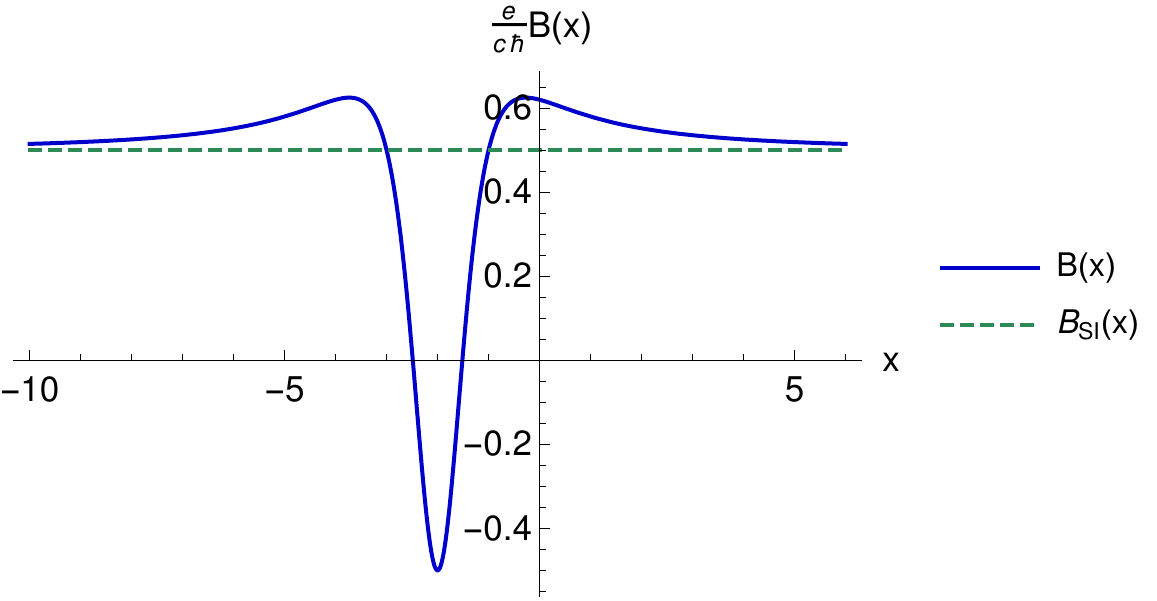}}
\caption{(a) Plot of the SUSY partner potentials $V_0$ and $V_2$, the first one being a shifted harmonic oscillator. The dashed line represents the shape-invariant SUSY partner potential of $V_0$. (b) Comparison between the new magnetic field $\mathcal{B}(x)$ of our approach and $\mathcal{B}_{SI}(x)$ for the shape-invariant case (the parameters have been fixed as $\omega=\kappa=1$).}
\label{Imagen3_1_1}
\end{center}
\end{figure}

\begin{figure}[ht] 
\begin{center}
\subfigure[]{\includegraphics[width=6.97cm, height=4.675cm]{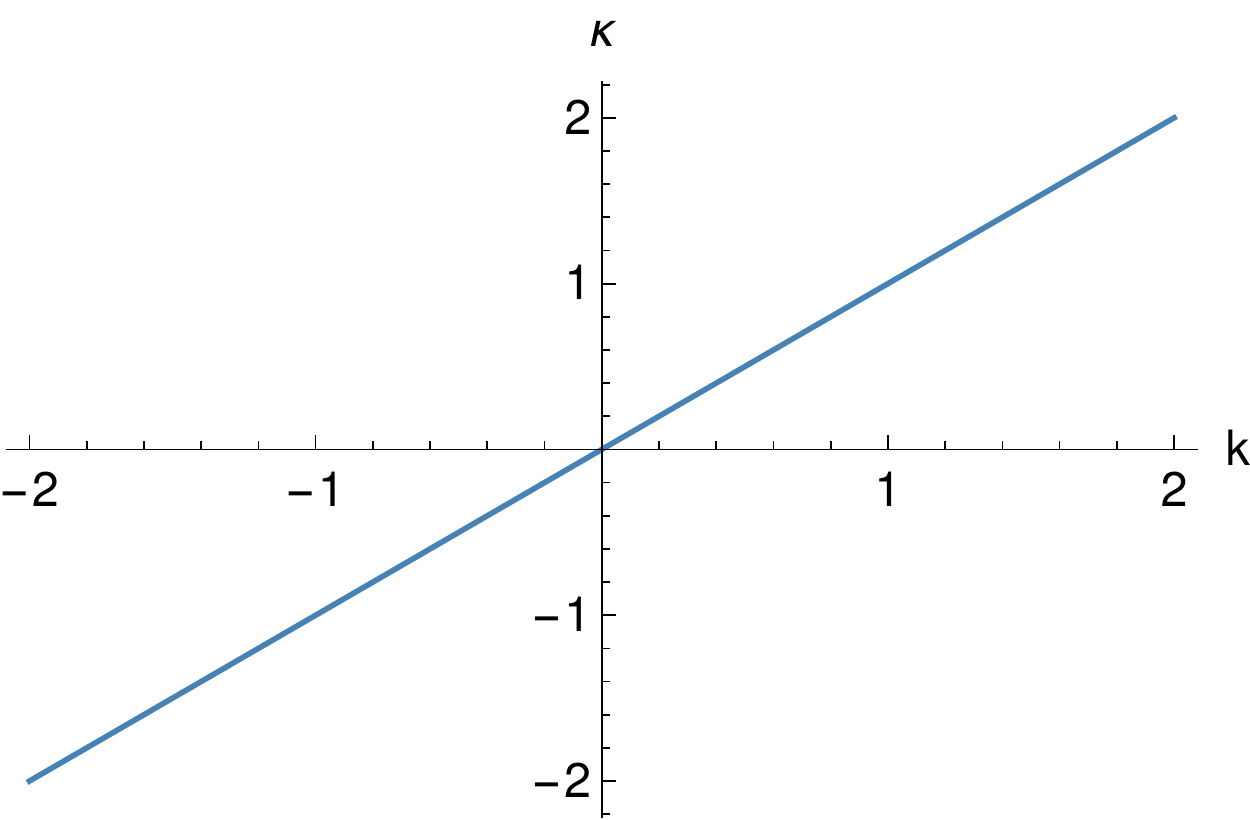}}
\subfigure[]{\includegraphics[width=6.97cm, height=4.675cm]{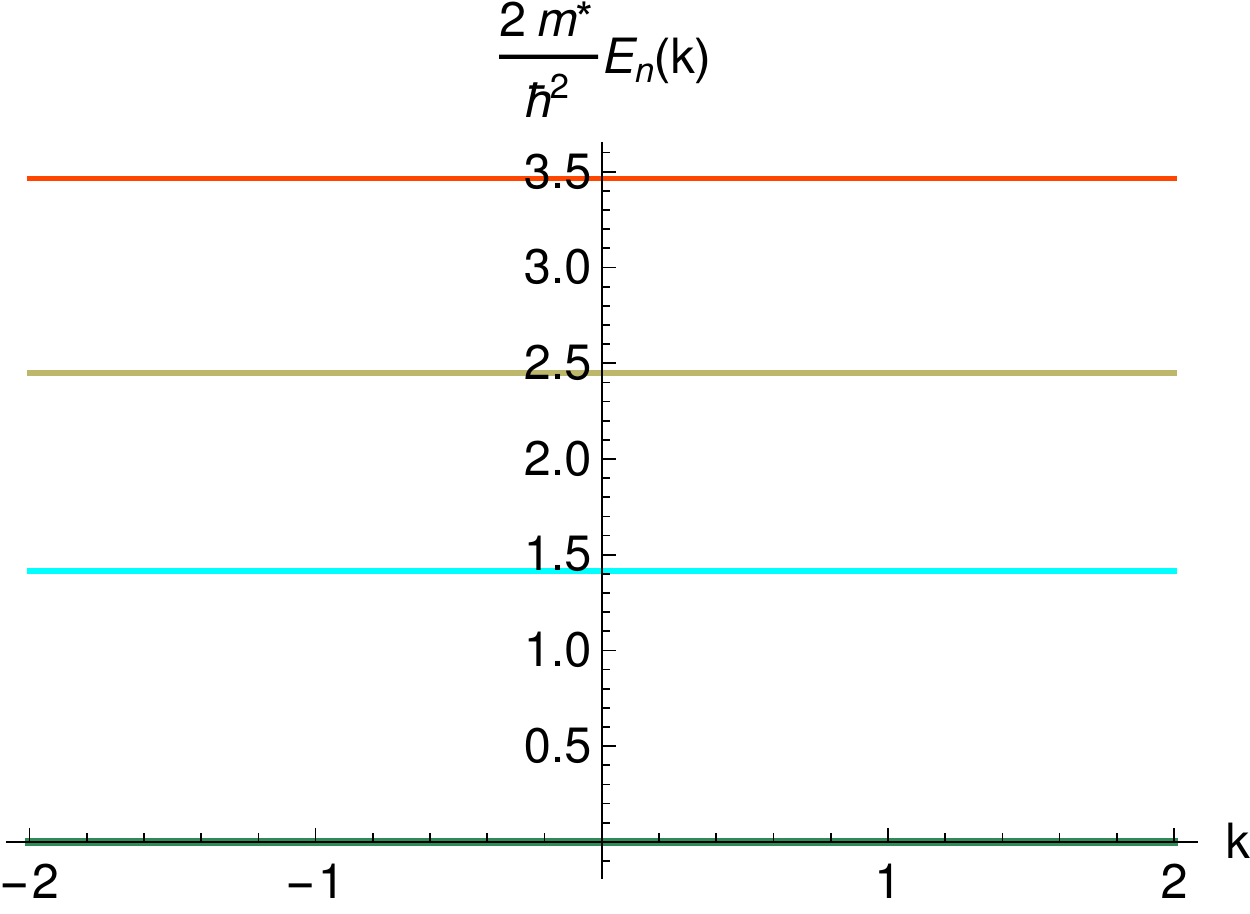}}
\caption{(a) Plot of $\kappa$ as function of the wave number $k$. (b) Eigenvalues ${E_n}$ as function of $k$ for $\omega=1$ and $j=1$. The two lowest energies, arising for $n=1,2$ and $n=0,3$ in equation (\ref{3.1.6}), are doubly degenerate.}
\label{Imagen3_1_2}
\end{center}
\end{figure}

\begin{figure}[ht] 
\begin{center}
\subfigure[]{\includegraphics[width=6.97cm, height=4.675cm]{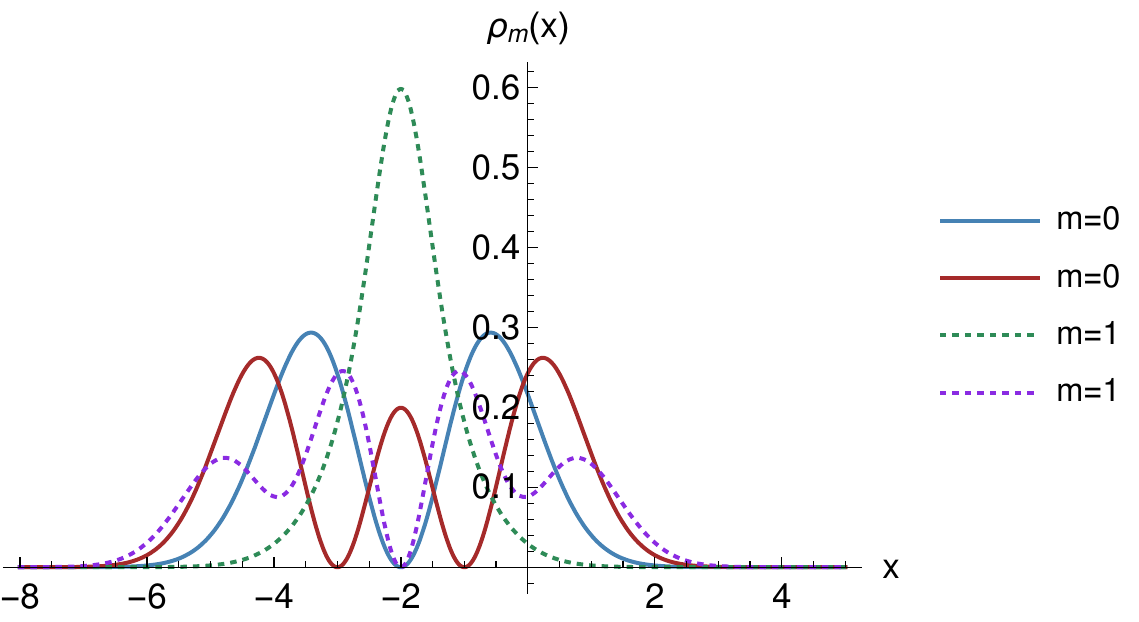}}
\subfigure[]{\includegraphics[width=6.97cm, height=4.675cm]{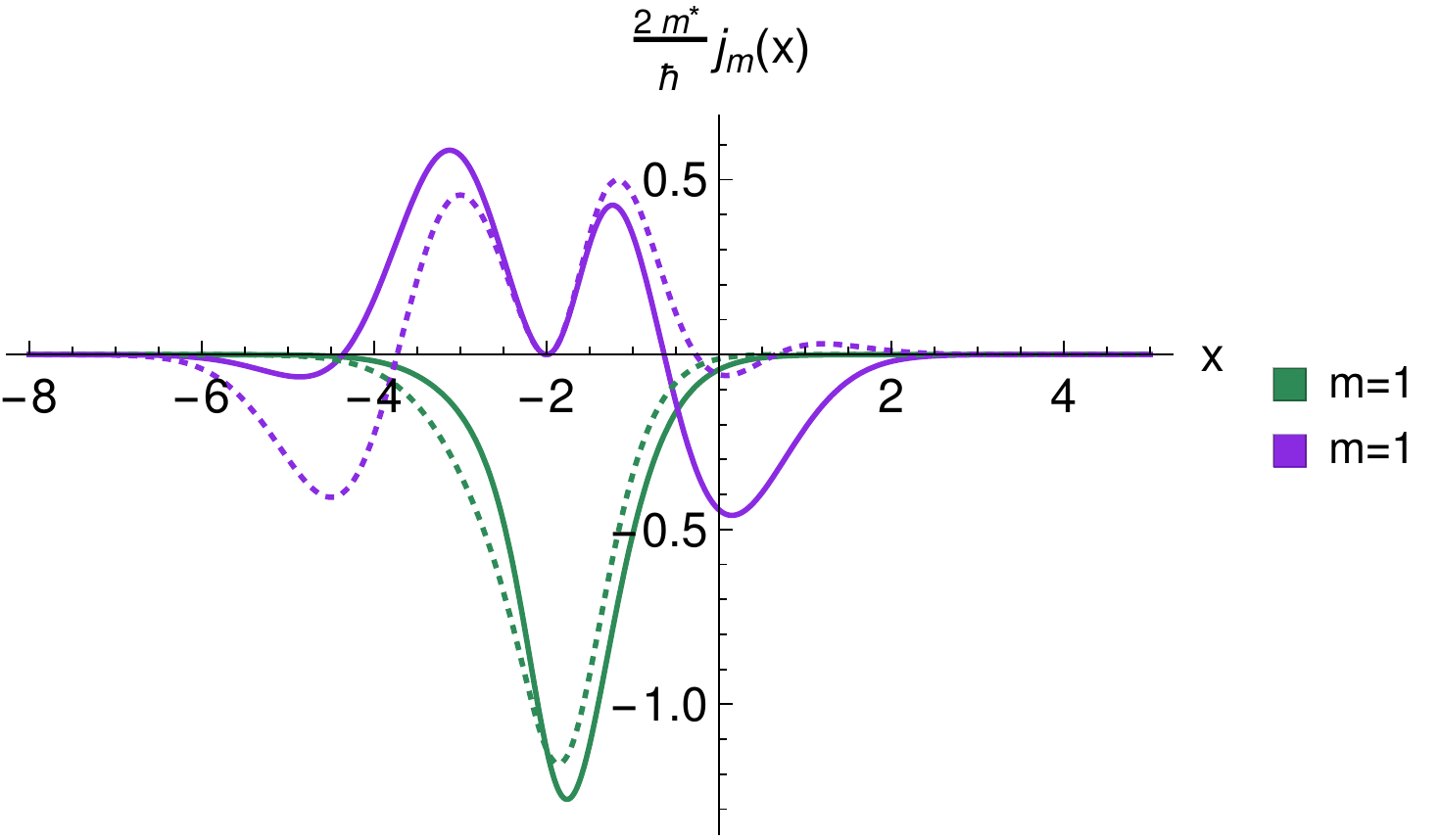}}
\caption{Plots of the probability density (a) and current densities $J_{x}$ (dotted lines) and $J_{y}$ (continuous lines) (b) for some energy eigenstates of bilayer graphene with $\omega=k=1$. The new label $m$ supplies now the standard ordering for the associated eigenvalues, which for $m=0$ and $m=1$ are doubly degenerate.}
\label{Imagen3_1_3}
\end{center}
\end{figure}

Figure~\ref{Imagen3_1_1} shows a plot of the SUSY partner potentials $V_0$, $V_2$ and a comparison of the magnetic field $\mathcal{B}(x)$ of equation (\ref{3.1.5}) with the field of the shape-invariant case. Figure~\ref{Imagen3_1_2} presents a plot of the parameter $\kappa$ as function of the wave number $k$ as well as the eigenenergies for electrons in bilayer graphene. Finally, the probability density and currents for several energy eigenstates are shown in Figure~\ref{Imagen3_1_3}. 

\subsection{Trigonometric Rosen-Morse potential}
The potential $V_0(x)$ is expressed now in terms of trigonometric functions and three real parameters $D,\kappa$ and $\alpha$ as follows
\begin{equation}
V_0(x)=D(D-\alpha)\csc^{2}(\alpha x)-2D\kappa\cot(\alpha x)-D^2+\kappa^2,
\label{3.2.1}
\end{equation}
where, once again, $\kappa$ will depend of the wavenumber $k$ (see Appendix A).
This is called Rosen-Morse potential of first kind or trigonometric Rosen-Morse potential, whose eigenfunctions $\psi_n^{(0)}$ and eigenvalues $\mathcal{E}_n^{(0)}$ are given by
\begin{align}
\psi_n^{(0)}(\zeta)&=c_n(-1)^{-\frac{s+n}{2}}(\zeta^2+1)^{-\frac{s+n}{2}}e^{a_n \, \text{arccot}(\zeta)}P_n^{(-s-n-ia_n,-s-n+ia_n)}(i\zeta), \label{3.2.2}
\\
\mathcal{E}_n^{(0)}&=\kappa^2-D^2+(D+n\alpha)^2-\frac{\kappa^2D^2}{(D+n\alpha)^2},\label{3.2.3}
\end{align}
with $c_n$ being a normalization constant, $s=D/\alpha$, $a_n=-\kappa D/\alpha (D+n\alpha)$, $\zeta=\cot(\alpha x)$, and $P^{(\alpha , \beta)}_n(\zeta)$ are the pseudo Jacobi polynomials. Note that an alternative expression for the eigenfunctions of the trigonometric Rosen-Morse potential in terms of real orthogonal polynomials can be found in \cite{Compean2005}. It is usual to find in the literature another form of the trigonometric Rosen-Morse potential, $V(x)=A(A-1)\csc^{2} x+2B\cot x-A^2+B^2/A^2$ \cite{Gangopadhyaya2018}. Exact analytic solutions for the bound states of $V(x)$ can be obtained for both $B\geq 0$ and $B<0$ \cite{Junker2019}. However, in order to recover the shape-invariant case explored in \cite{fgo20}, we will choose the initial potential as in equation~\eqref{3.2.1}, which is reached from $V(x)$ by taking $A=D$ and $B=-\kappa D$.

We have calculated the function $\eta$ from the bound states associated to the factorization energies $\epsilon_1=\mathcal{E}^{(0)}_{2}$, $\epsilon_2=\mathcal{E}^{(0)}_1$ (see also \cite{df11}). Then, the SUSY partner potential $V_2(x)$ of $V_0(x)$ and the magnetic field leading to that pair are given by 
\begin{align}
& V_2(x)=V_{SI}(x)+8\alpha^2\left[\kappa^2+(D+2\alpha)^2\right]\frac{\left(1+\zeta^2\right)(D+\alpha)(2D+3\alpha)\zeta^2-2\kappa (2D+3\alpha)\zeta+2\kappa^2-\alpha(D+\alpha)}{\left[(D+2\alpha)(2D+3\alpha)\zeta^2-2\kappa (2D+3\alpha)\zeta+2\kappa^2+\alpha(D+2\alpha)\right]^2}, \\
& \mathcal{B}(x) = \mathcal{B}_{SI}(x)\left[1+4\alpha\frac{\kappa^2+(D+2\alpha)^2}{2D+\alpha}\frac{(D+\alpha)(2D+3\alpha)\zeta^2-2\kappa (2D+3\alpha)\zeta+2\kappa^2-\alpha(D+\alpha)}{\left[(D+2\alpha)(2D+3\alpha)\zeta^2-2\kappa (2D+3\alpha)\zeta+2\kappa^2+\alpha(D+2\alpha)\right]^2}\right].
\label{3.2.4}
\end{align}

\begin{figure}[ht] 
\begin{center}
\subfigure[]{\includegraphics[width=8.2cm, height=5.5cm]{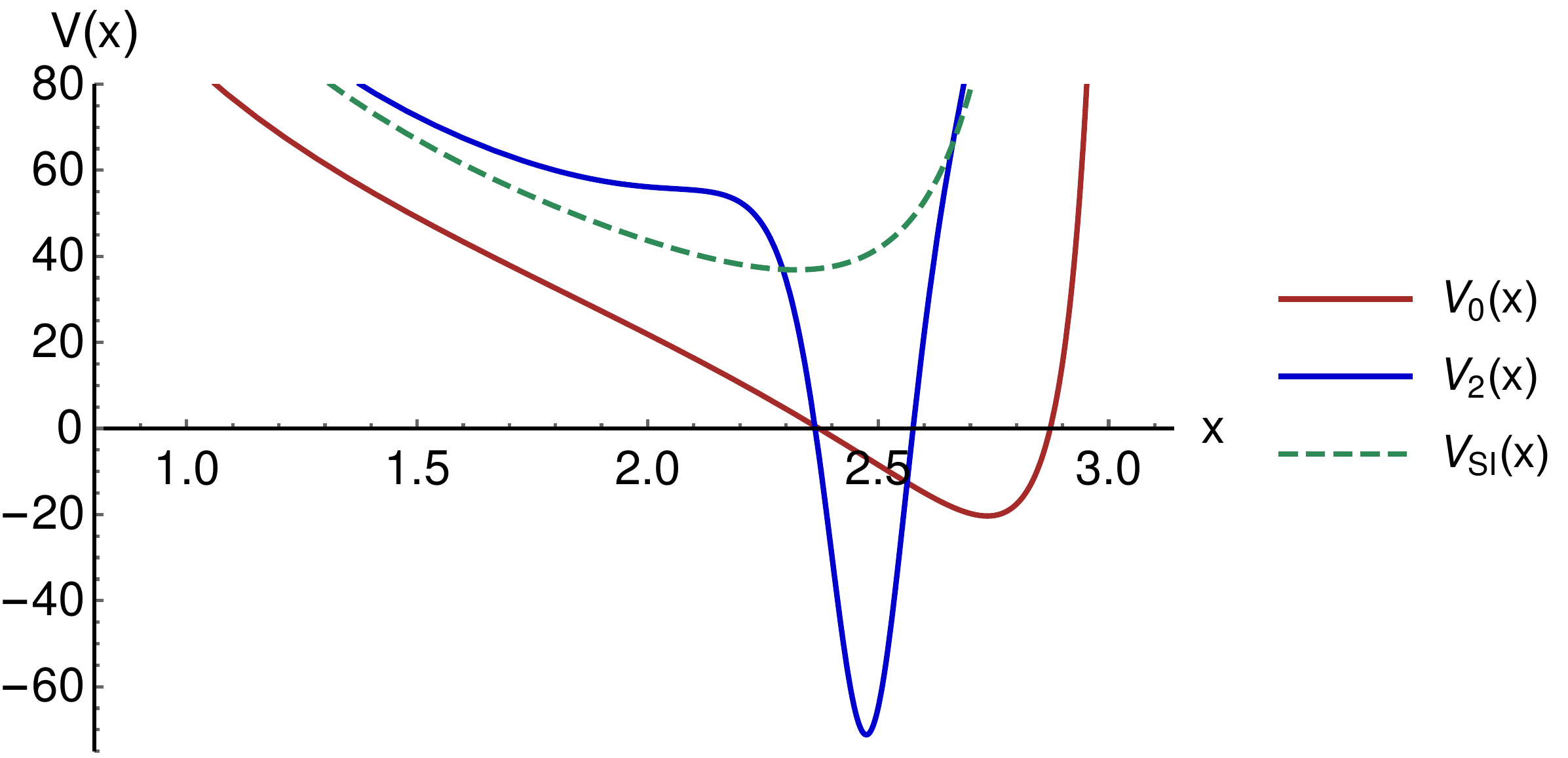}}
\subfigure[]{\includegraphics[width=8.2cm, height=5.5cm]{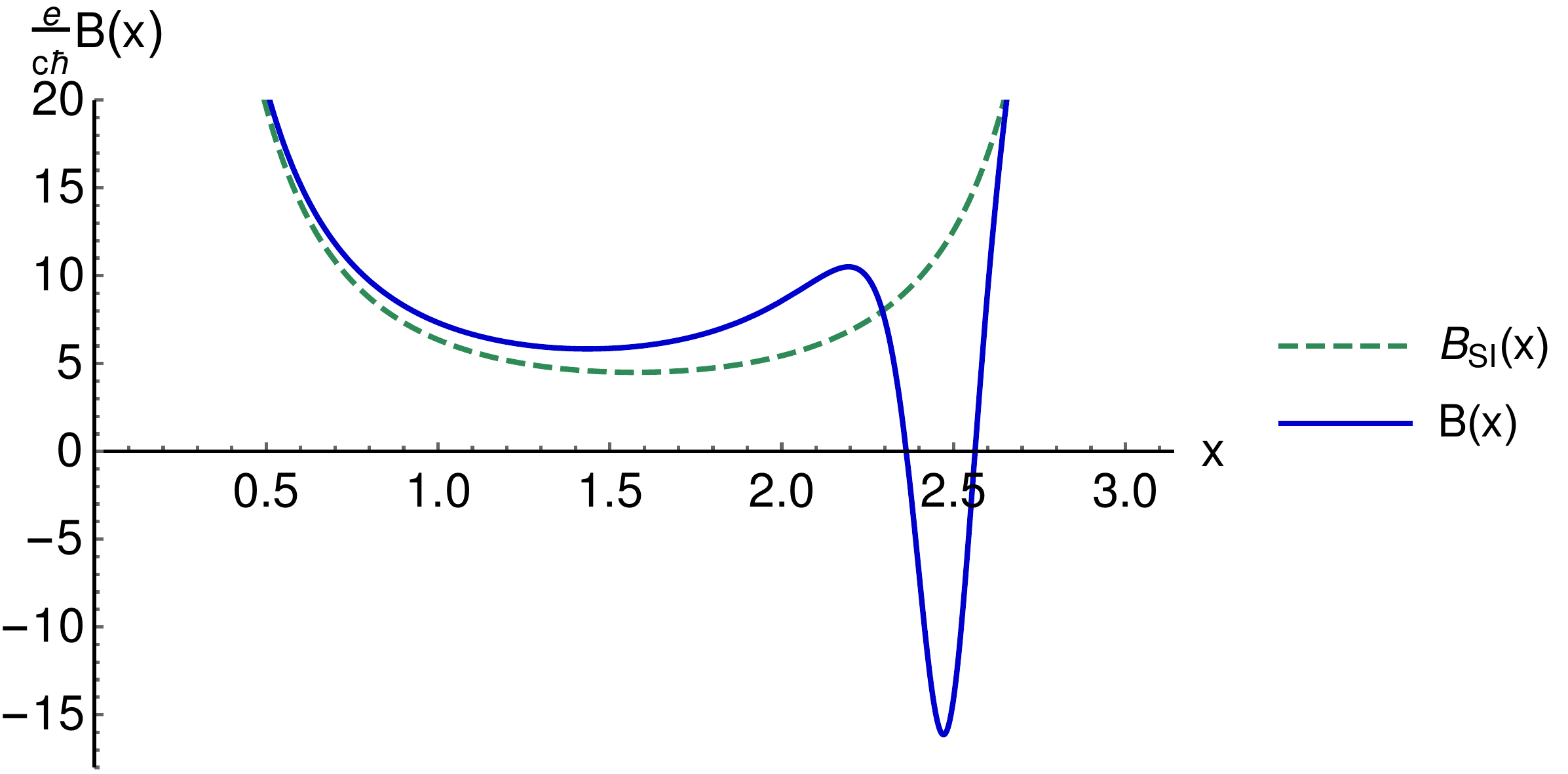}}
\caption{(a) Plot of the potentials $V_0(x)$, $V_2(x)$ and the corresponding SUSY partner potential in the shape-invariant case (dashed line) for the trigonometric Rosen-Morse potential. (b) The new magnetic field $\mathcal{B}(x)$ (continuous line) as compared with $\mathcal{B}_{SI}(x)$ (dashed line). The parameters were taken as $D=4, \alpha=1$ and $\kappa=-7$.}\label{F-3-2-1}
\end{center}
\end{figure}

We can see that this magnetic field differs from the one leading to the shape-invariant case, $\mathcal{B}_{SI}(x)=\frac{c\hbar\alpha}{2e}(2D+\alpha)\left(1+\zeta^2\right)$, by a rational factor while $V_{SI}(x)=(D+\alpha)(D+2\alpha)(1+\zeta^2)-2D\kappa\zeta-D^2+\kappa^2$. In Figure~\ref{F-3-2-1} we show plots of the potentials $V_0(x), \ V_2(x)$ and the magnetic fields $\mathcal{B}(x)$, $\mathcal{B}_{SI}(x)$.

It is important to notice that the wavenumber $k$ is now a non-trivial function of the parameter $\kappa$:
\begin{equation}
k=\frac{\kappa(2D+3 \alpha)}{2(D+\alpha) (D+2 \alpha)}\cdot\frac{(D+\alpha) (D+2 \alpha)^2-D \kappa ^2}{(D+\alpha) (D+2 \alpha)-\kappa ^2}.
\label{3.2.5}
\end{equation}

For our original problem, the electron in bilayer graphene under the magnetic field \eqref{3.2.4}, the resulting positive energy eigenvalues (the negative ones correspond to holes) are given by equation~(\ref{3.2}) for $j=1$, where
\begin{eqnarray}
&& \Delta_{n,1}=D^2 \kappa^2 \left(\frac{1}{(D+\alpha)^2}-\frac{1}{(D+\alpha n)^2}\right)+(n-1)\left[2D+(n+1)\alpha\right]\alpha,
\label{3.2.6} \\
&& \Delta_{n,2}=D^2 \kappa^2 \left(\frac{1}{(D+2 \alpha)^2}-\frac{1}{(D+\alpha n)^2}\right)+(n-2)\left[2D+(n+2)\alpha\right]\alpha.
\label{3.2.7}
\end{eqnarray}

\begin{figure}[ht] 
\begin{center}
\subfigure[]{\includegraphics[width=8.2cm, height=5.5cm]{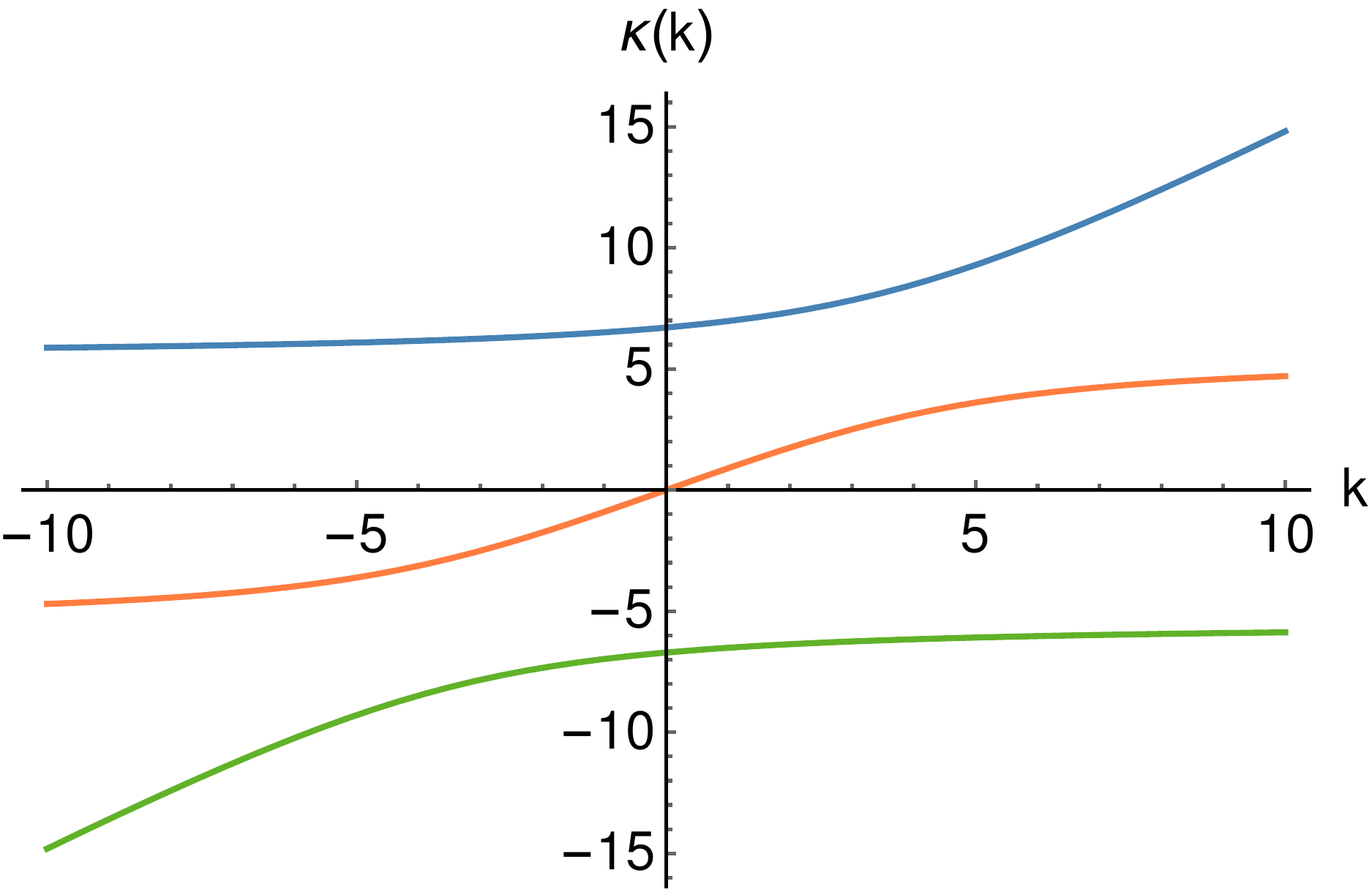}}
\subfigure[]{\includegraphics[width=8.2cm, height=5.5cm]{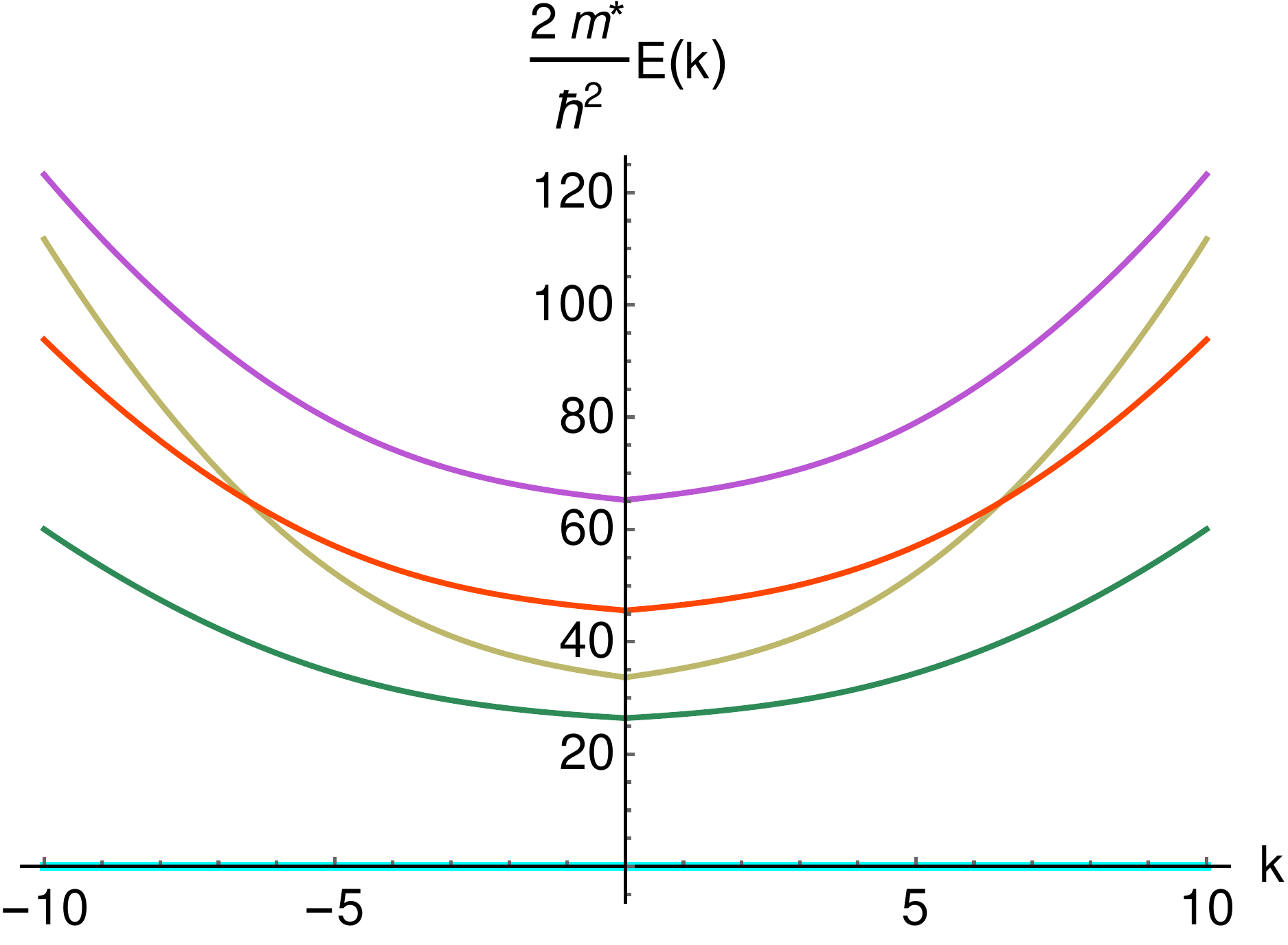}}
\caption{(a) The three possible forms of $\kappa(k)$ for the trigonometric Rosen-Morse potential. (b) Electron eigenenergies as functions of $k$ (the ground state energy is doubly degenerate). The parameters were taken as $D=4, \alpha=1$.}\label{F-3-2-2}
\end{center}
\end{figure}

\begin{figure}[ht]
\begin{center}
\subfigure[]{\includegraphics[width=8.2cm, height=5.5cm]{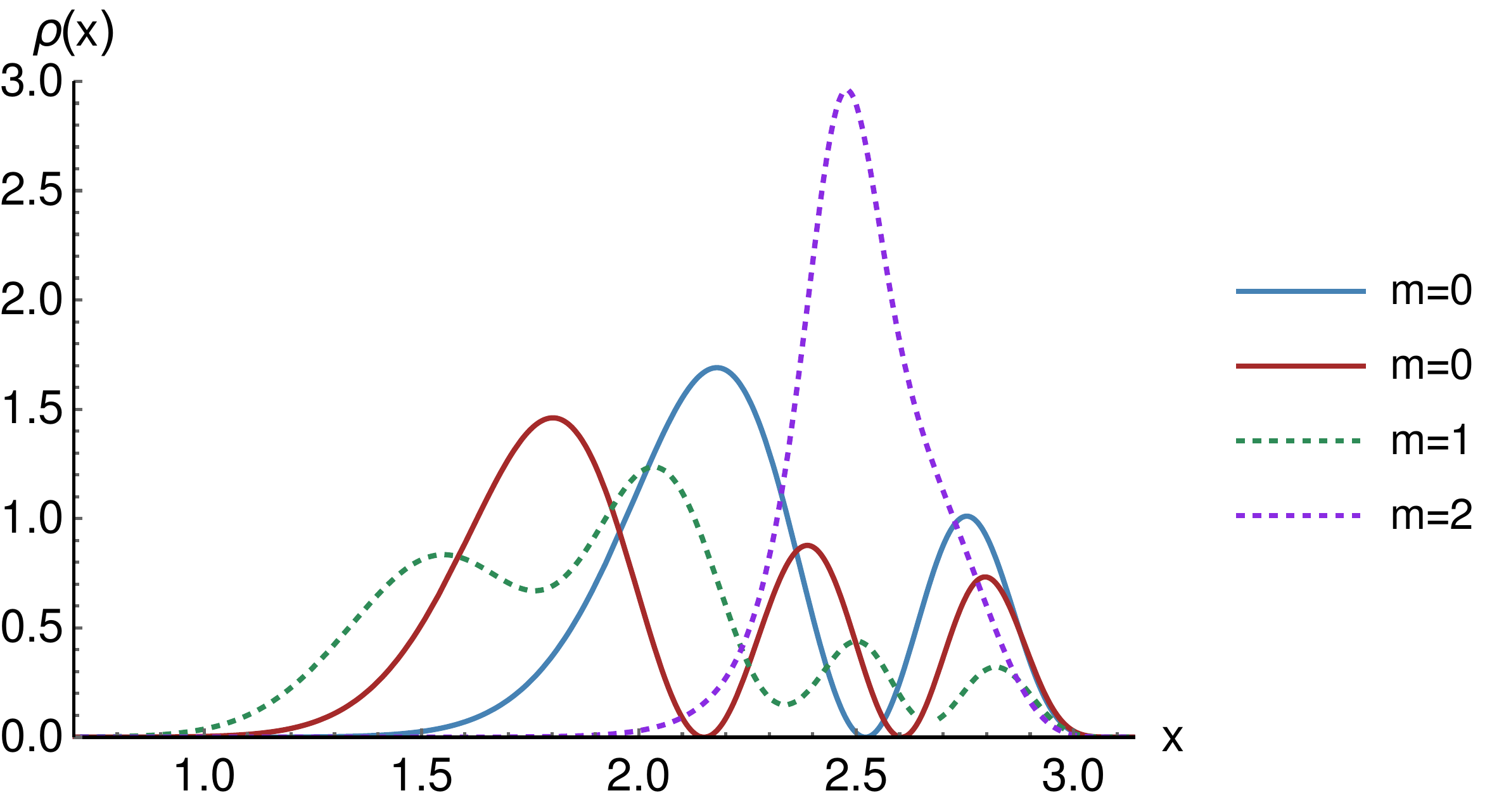}}
\subfigure[]{\includegraphics[width=8.2cm, height=5.5cm]{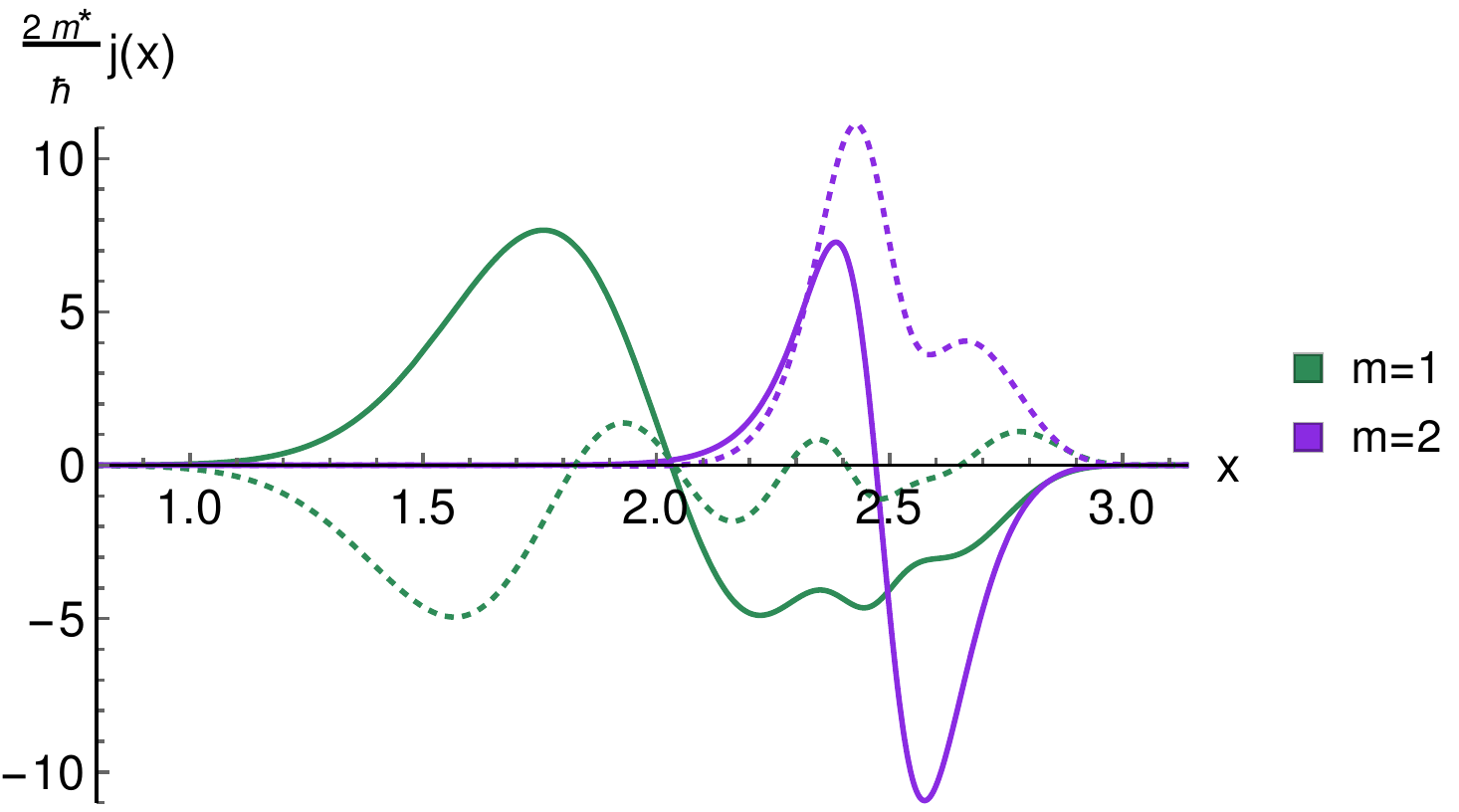}}
\caption{Plot of the probability (a) and current densities (b) for the trigonometric Rosen-Morse potential with $D=4, \alpha=1$ and $\kappa=-7$. For this parameters choice the standard ordering is shown. Note that the ground state energy is doubly degenerate (see the two continuous lines in (a)). In (b) the current density $J_{x}$ corresponds to the dotted lines while $J_{y}$ does for the continuous lines.}\label{F-3-2-3}
\end{center}
\end{figure}

Although such electron energies depend on $\kappa$, this does not impose any further restriction on them. However, the standard ordering of these energy levels depends on the parameters $D$, $\kappa$ and $\alpha$. Note that the ground state energy is always doubly degenerate. Moreover, when solving equation \eqref{3.2.5} three solutions for $\kappa(k)$ will be obtained, which are plotted in Figure \ref{F-3-2-2}. If we remember that for low energies the electron energy must be quadratic in the momentum \cite{McCann2013}, we can determine which ones of these solutions lead to the appropriate electron energies in bilayer graphene. In Figure \ref{F-3-2-3} the probability and current densities for the eigenvectors associated to the lowest electron energies are shown.

\subsection{Hyperbolic Rosen-Morse potential}
\begin{figure}[ht] 
\begin{center}
\subfigure[]{\includegraphics[width=6.97cm, height=4.675cm]{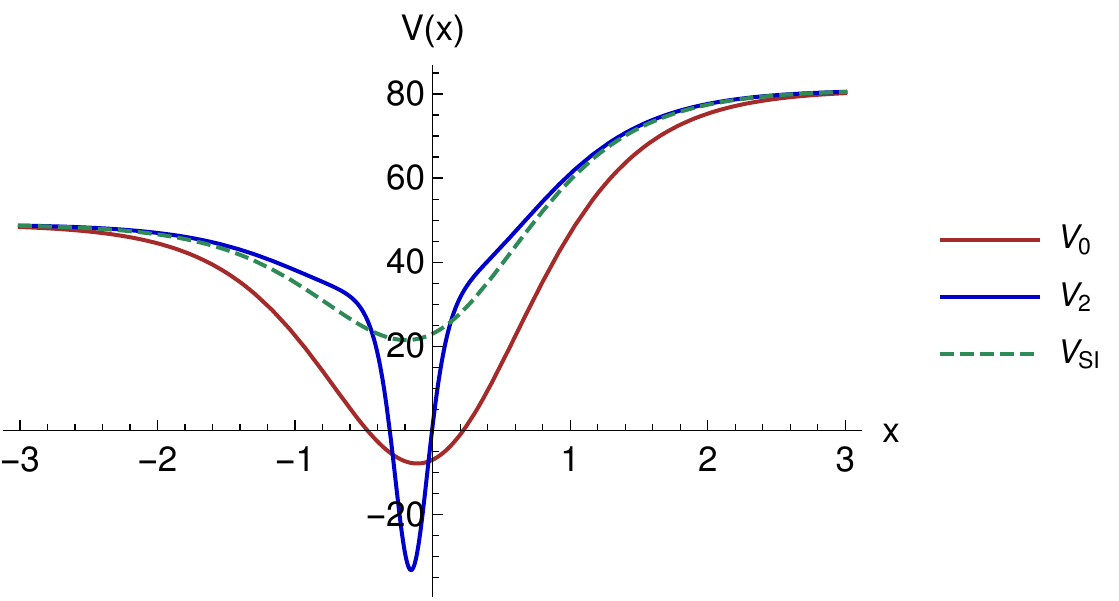}}
\subfigure[]{\includegraphics[width=6.97cm, height=4.675cm]{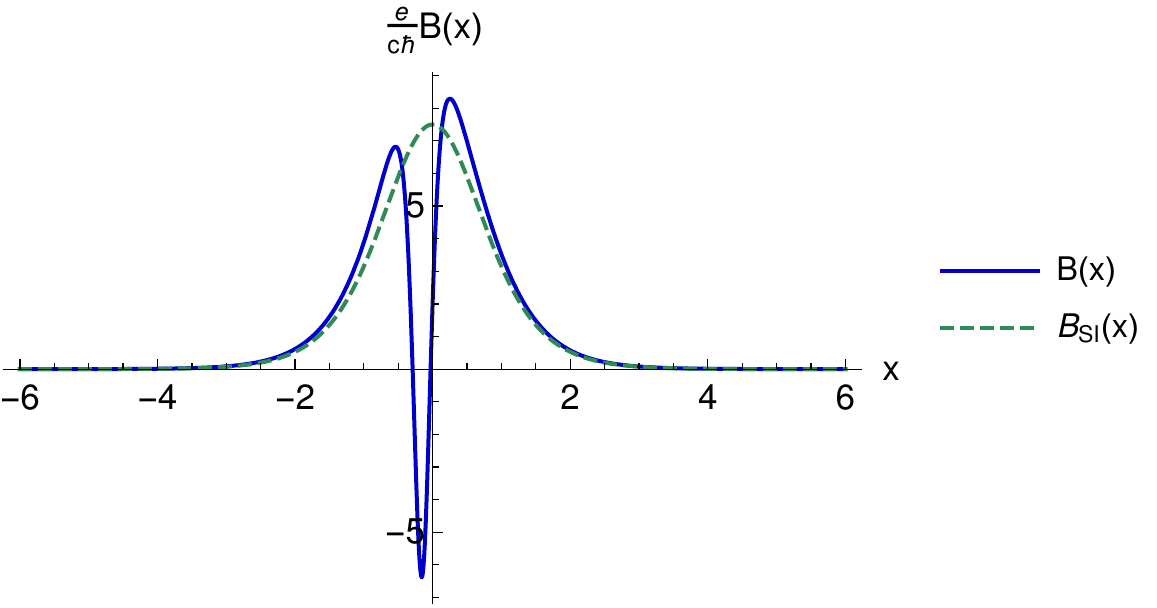}}
\caption{(a) Plot of $V_0(x)$ (Rosen-Morse II potential), as well as its non shape-invariant and shape-invariant SUSY partners $V_2(x)$ and $V_{SI}(x)$. (b) Comparison between the magnetic field $\mathcal{B}(x)$ and the shape-invariant limit $\mathcal{B}_{SI}(x)$. The parameters have been taken as $D=8$ and $\alpha=\kappa=1$.}
\label{Imagen3_3_1}
\end{center}
\end{figure}

In this case we have that
\begin{equation}
V_{0}={D}^{2}+\kappa^{2}-{D}({D}+\alpha)\textrm{sech}^{2}(\alpha x)+2\kappa{D}\textrm{tanh}(\alpha x),
\label{3.3.1}
\end{equation}
which is called either Rosen-Morse II potential or hyperbolic Rosen-Morse potential, where $D$, $\kappa$ and $\alpha$ are real parameters. As in the previous cases, we assume that the parameter $\kappa$ will depend on the wavenumber $k$. The corresponding eigenfunctions $\psi_n^{(0)}$ and eigenvalues $\mathcal{E}_n^{(0)}$ are given by
\begin{eqnarray}
&& \psi_n^{(0)}(\zeta) = c_n\left(1-\zeta\right)^{\frac{s-n+a_n}{2}}\left(1+\zeta\right)^{\frac{s-n-a_n}{2}}\textrm{P}_n^{(s-n+a_n,s-n-a_n)}(\zeta),
\label{3.3.3}\\
&& \mathcal{E}^{(0)}_{n} = {D}^{2}+\kappa^{2}-\left({D}-n\alpha\right)^{2}-\frac{\kappa^{2}{D}^{2}}{\left({D}-n\alpha\right)^{2}}, \quad n=0,1,\dots,N,
\label{3.3.2}
\end{eqnarray}
where $N$ defines the maximum number of excited states, $c_n$ is a normalization factor, $\zeta=\textrm{tanh}(\alpha x)$, $s=\frac{{D}}{\alpha}$, $a_n=\frac{{D}\kappa}{\alpha({D}-n\alpha)}$ and
$\textrm{P}_n^{(a,b)}(\zeta)$ are the Jacobi polynomials. Unlike the two previous cases, here we have a finite discrete spectrum. In order to determine the number of bound states fulfilling the square-integrability condition, the following relationship must be valid for all the parameter values:
\begin{equation}
|\kappa|<\frac{(D-\alpha  n)^2}{D}.
\label{3.3.4}
\end{equation}
By taking as seed solutions the first two excited states with their corresponding eigenvalues (we are assuming that $V_0$ has at least three bound states), we obtain straightforwardly the SUSY partner potential $V_2$ and the magnetic field for this case:
\begin{align}
V_2(x)&=V_{SI}(\zeta)+8 \alpha^2\left[(D-2\alpha)^2-\kappa^2\right]\frac{(1-\zeta^2)\left[2(D\zeta+\kappa)^2-(5D\zeta^2+6\kappa \zeta +D)\alpha+(3\zeta^2 +1)\alpha^2\right]}{\left[2 \kappa^2 +(D-2\alpha)\left(\zeta^2(2D-3\alpha)+\alpha\right)+2\kappa\zeta (2D -3 \alpha)\right]^2},\\
\mathcal{B}(x)=&\mathcal{B}_{SI}(\zeta)
\left[1+\frac{4\alpha\left[(D-2\alpha)^2-\kappa^2\right] \left[2(D\zeta+\kappa)^2-(5D\zeta^2+6\kappa \zeta +D)\alpha+(3\zeta^2 +1)\alpha^2\right]}{(2 D-\alpha ) \left[2 \kappa^2 +(D-2\alpha)\left(\zeta^2(2D-3\alpha)+\alpha\right)+2\kappa\zeta (2D -3 \alpha)\right]^2}\right].
\label{3.3.5}
\end{align}
It can be seen how different our new SUSY partner potential and magnetic field are compared with those of the shape-invariant case, $\mathcal{B}_{SI}(x)=\frac{c\hbar\alpha}{2e}(2D-\alpha)(1-\zeta^2)$ and $V_{SI}(\zeta)={D}^{2}+\kappa^{2}-({D}-\alpha)({D}-2\alpha)(1-\zeta^2)+2\kappa{D}\zeta$. In Figure~\ref{Imagen3_3_1}~ it is shown the SUSY partner potentials $V_0$, $V_2$ and the magnetic field $\mathcal{B}(x)$ of equation (\ref{3.3.5}) as compared with the field in the shape-invariant case.

On the other hand, according to equation (\ref{2.7}) the relationship between $\kappa$ and $k$ becomes now
\begin{equation}
k=\frac{(2D-3\alpha)\left[D\kappa^3+\kappa(D^3-5\alpha D^2+8\alpha^2 D-4\alpha^3)\right]}{2(D-2\alpha)(D-\alpha)\left[\kappa^2+(D^2-3\alpha D+2\alpha^2)\right]}.
\label{3.3.6}
\end{equation}
For the magnetic field of equation (\ref{3.3.5}) the eigenvalues for the electron in bilayer graphene are given by equation~(\ref{3.2}) with $j=1$, where
\begin{eqnarray}
&& \Delta_{n,1}=D^2 \kappa^2 \left(\frac{1}{(D-\alpha)^2}-\frac{1}{(D-\alpha  n)^2}\right)+\alpha (n-1)[2D-\alpha(n+1)],
\label{3.3.7}
\\
&& \Delta_{n,2}=D^2\kappa^2 \left(\frac{1}{(D-2\alpha  )^2}-\frac{1}{(D-\alpha  n)^2}\right)+\alpha (n-2)[2D-\alpha(n+2)].
\label{3.3.8}
\end{eqnarray}

\begin{figure}[ht] 
\begin{center}
\subfigure[]{\includegraphics[width=6.97cm, height=4.675cm]{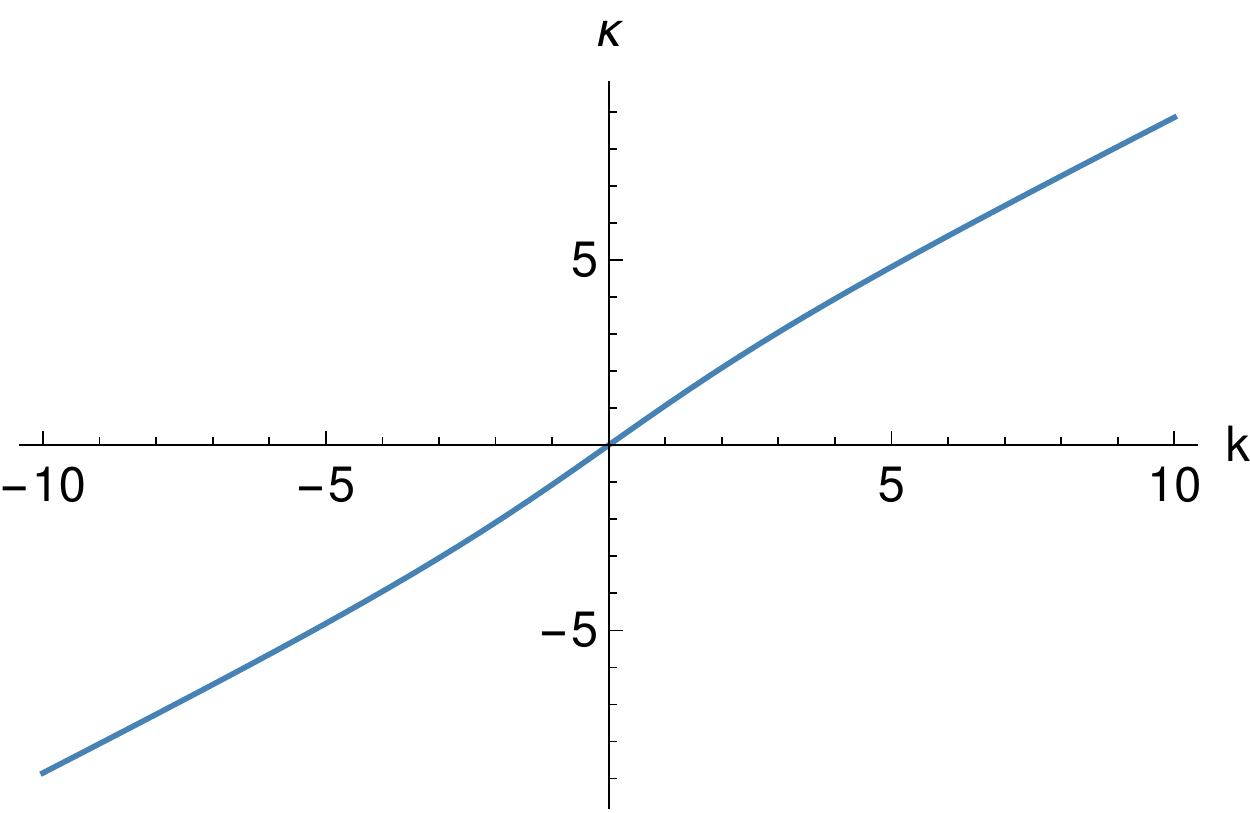}}
\subfigure[]{\includegraphics[width=6.97cm, height=4.675cm]{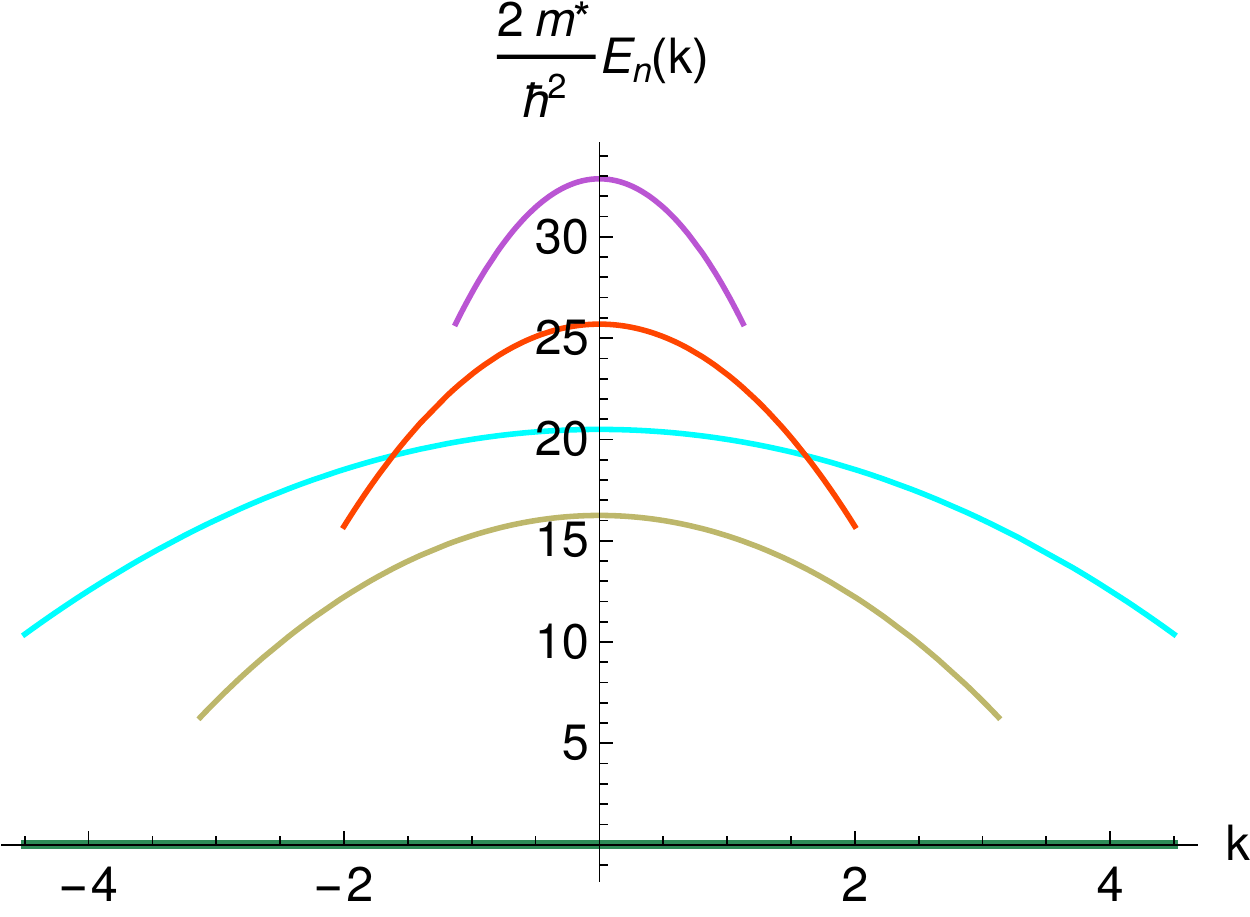}}
\caption{(a) Plot of $\kappa$ versus $k$ for the Rosen-Morse II potential with $D=8$ and $\alpha=1$, which is not a linear function. (b) Eigenenergies $E_{n}$ as functions of $k$ for the same $D$ and $\alpha$.}
\label{Imagen3_3_2}
\end{center}
\end{figure}

Let us stress the importance of equation~(\ref{3.3.6}), since together with equations (\ref{3.3.7}) and (\ref{3.3.8}) at the end indicate that the eigenenergies are functions of $k$. Note as well that the standard ordering for $E_n$ will depend on the election of all the parameters $D$, $\alpha$ and $k$ (see~Figure~\ref{Imagen3_3_2}).

\begin{figure}[ht] 
\begin{center}
\subfigure[]{\includegraphics[width=6.97cm, height=4.675cm]{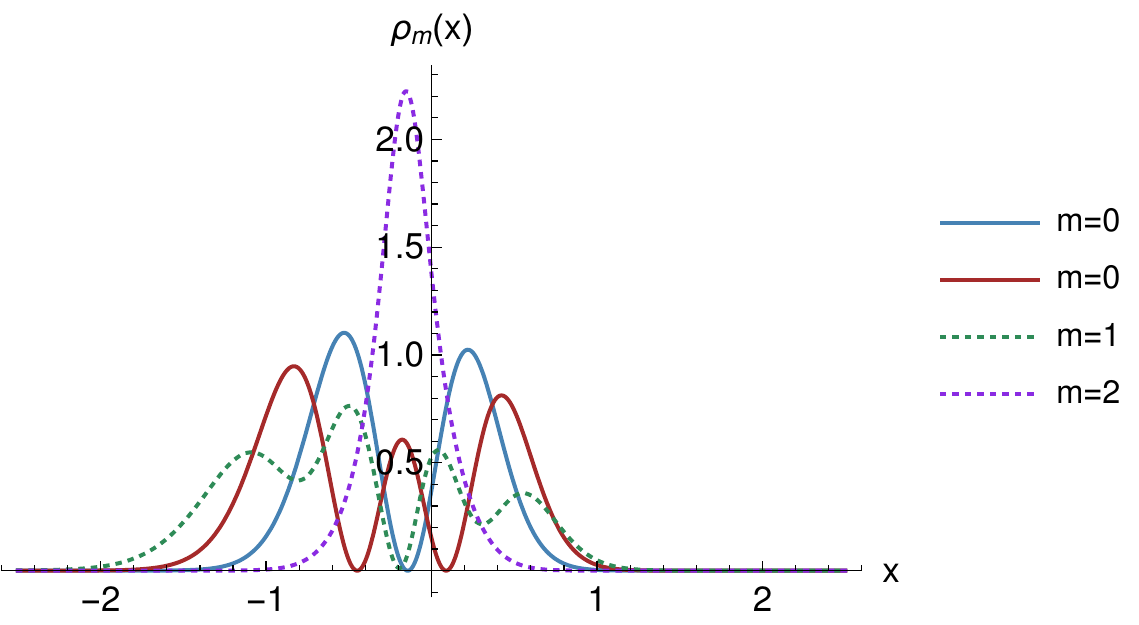}}
\subfigure[]{\includegraphics[width=6.97cm, height=4.675cm]{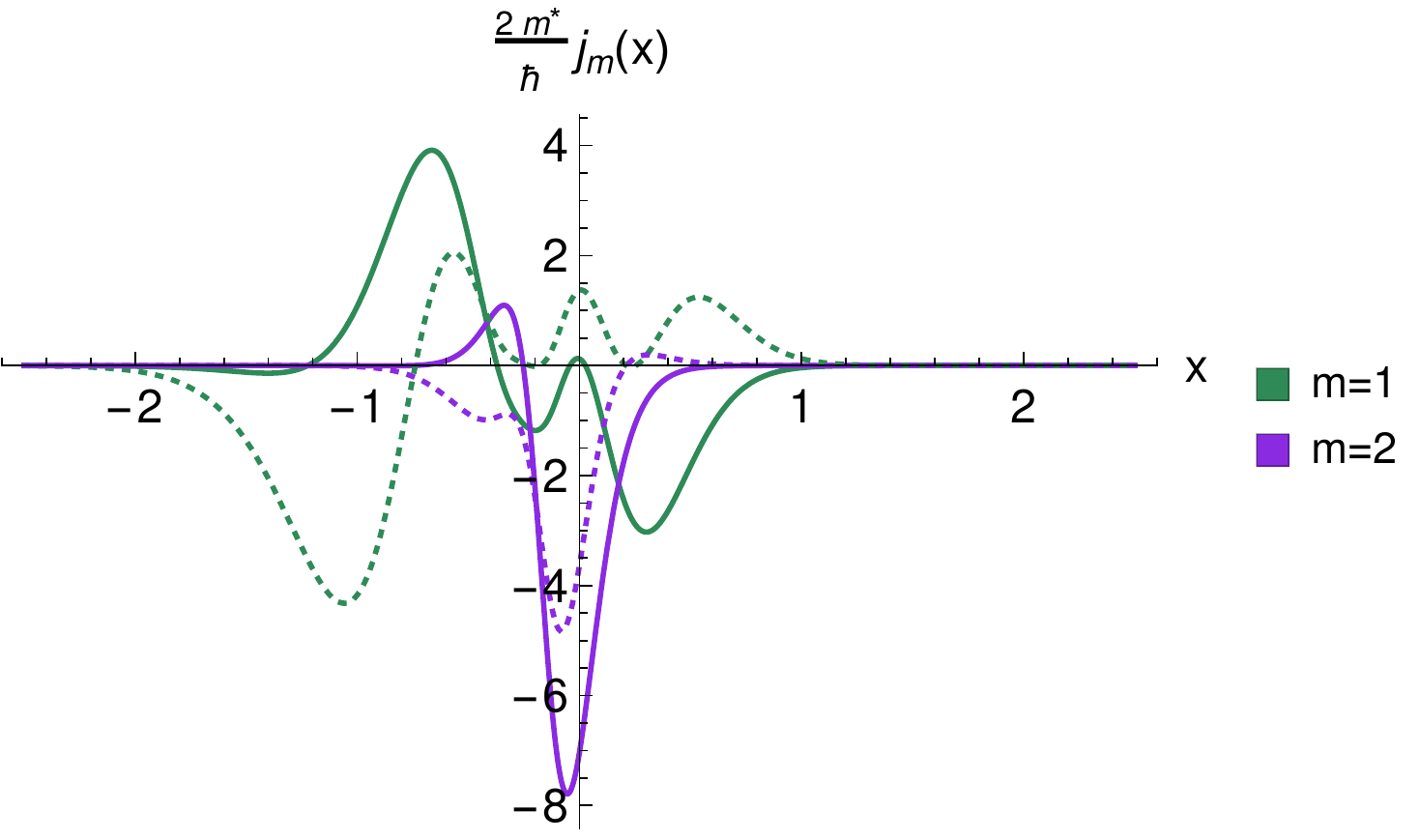}}
\caption{Plot of the probability density (a) and current density $J_{x}$ (dotted lines) and $J_{y}$ (continuous lines)(b) for some eigenstates of bilayer graphene when $V_0$ is the Rosen-Morse II potential with $D=8$, $\alpha=\kappa=1$. The ordering defined by the index $m$ is the standard one for this parameters choice.}
\label{Imagen3_3_3}
\end{center}
\end{figure}

A plot of the parameter $\kappa$ as function of the wavenumber $k$, as well as the eigenenergies, is given in Figure~\ref{Imagen3_3_2}. We are showing also the probability density and current in Figure~\ref{Imagen3_3_3} where, unlike Figure~\ref{Imagen3_3_2}, we have taken a fixed value of $k$ (thus of $\kappa$) to be able to identify the standard ordering of the non-degenerate excited states. 

\section{The confluent algorithm}
The characteristic fact of this algorithm is that we just need to choose one seed solution $u^{(0)}(x)$, associated to the only factorization energy $\epsilon_1=\epsilon_2=\epsilon$ \cite{Salinas2003,Salinas2005} (see also \cite{Barnana2020,mnr00,fs11,bff12,cjp15,gq15,cs15a,cs15b,be16,cs17,sy18}). As in the previous case, the SUSY transformation is defined by a function $\eta (x)$ such that
\begin{equation}
\eta(x)=-\frac{\mathrm{w}'(x)}{\mathrm{w}(x)},
\label{4.2}
\end{equation} 
where 
\begin{equation}
\mathrm{w}(x)=\mathrm{w}_0-\int_{x_0}^{x}\big[u^{(0)}(y)\big]^2dy.
\label{4.1}
\end{equation}
Let us recall that $u^{(0)}(x)$ must be in the kernel of $L_2^{-}$ and should be a formal eigenfunction of $H_0$ associated to $\epsilon$, while $x_0$ is a fixed point in the domain of $V_0(x)$. Furthermore, the constant $\mathrm{w}_0$ can be adjusted to guarantee that $\mathrm{w}(x)$ will be nodeless. Thus, although in the confluent case at the beginning we just supply the factorization energy and its corresponding seed solution, at the end we will have a one-parameter family of magnetic fields $\mathcal{B}(x)=\mathcal{B}(x; \mathrm{w}_0)$ and SUSY partner potentials, $V_2(x)=V_2(x;\mathrm{w}_0)$ parametrized by $\mathrm{w}_0$. As we established previously, the seed solution will be taken as the eigenstate of $H_0$ associated to $\mathcal{E}^{(0)}_j$, $u^{(0)}(x)=\psi_j^{(0)}(x)$. Also, it is convenient to select $x_0$ as the left end of the domain of $V_0(x)$. Due to this, $\mathrm{w}(x)$ will be nodeless either for $\mathrm{w}_0\geq 1$ or 
$\mathrm{w}_0\leq 0$. In particular, if $\mathrm{w}_0>1$ or $\mathrm{w}_0<0$ the transformation will produce a Hamiltonian $H_2$ which is isospectral to $H_0$, i.e., $\epsilon$ belongs to the spectrum of $H_2$. On the other hand, if $\mathrm{w}_0$ would be either $0$ or $1$, we will get that $\epsilon$ is not an eigenvalue of $H_2$, so that $H_0$ and $H_2$ become almost isospectral in this limit. Analogously to the previous case (see equation~\eqref{2.15}), the eigenfunctions $\psi_n^{(2)}(x)$ of $H_2$ for $n\neq j$ are given by
\begin{equation}
\psi_n^{(2)}(x)=\frac{L_2^{-}\psi_n^{(0)}(x)}{\left|\mathcal{E}^{(0)}_n-\mathcal{E}^{(0)}_j\right|},\quad n=0,1,...,j-1,j+1,\dots,
\label{4.3}
\end{equation}
with $\mathcal{E}_j^{(0)}$ being the chosen factorization energy. The normalizability of the formal eigenfunction of $H_2$ associated to $\epsilon=\mathcal{E}_j^{(0)}$, $\psi_j^{(2)}(x)\propto\frac{\psi_j^{(0)}(x)}{\mathrm{w}(x)}$, has to be checked in order to see weather $\epsilon$ belongs or not to Sp($H_2$).
 
Going back to our initial problem, the energy eigenvalues for electrons in bilayer graphene in the isospectral case turn out to be 
\begin{equation}
E_n=\frac{\hbar^2}{2m^*}|\Delta_{n,j}|,
\label{4.4}
\end{equation}
whose eigenfunctions are
\begin{equation}
\Psi_{n}(x,y) =\frac{e^{iky}}{\sqrt{2}}
\begin{pmatrix}
\psi_{n}^{(2)}(x)\\
\psi_n^{(0)}(x)
\end{pmatrix},
\qquad\text{for}\ n=0,1,\dots
\label{4.5}
\end{equation}
Note that these eigenfunctions are valid only for $w_0<0$ or $w_0>1$. When either $\mathrm{w}_0=0$ or $\mathrm{w}_0=1$, equation~(\ref{4.5}) is still valid for $n\neq j$, but for $n=j$ the first entry in $\Psi_j$ must be set equal to zero and we must drop the factor $1/\sqrt{2}$. 

As in the previous case, in general the index $n$ of the auxiliary Schr\"odinger equations does not supply the standard ordering of the bilayer graphene energies. It is interesting to see as well that the ground state energy in the confluent case is no longer degenerate, as in the non-confluent case or in the shape-invariant situation. In the next examples we will calculate the integral in equation (\ref{4.1}) for an arbitrary $j$, but for other calculations and plots we will take $j=0$.

\subsection{Shifted harmonic oscillator}
Let us take $V_0(x)$ as the harmonic oscillator potential of equation \eqref{3.1.1}, whose energy eigenfunctions and eigenvalues are given by equations \eqref{3.1.3} and \eqref{3.1.2}, respectively. By choosing the seed solution and $x_0$ as mentioned at the beginning of this section, the integral in equation~\eqref{4.1} can be written as follows
\begin{equation} \label{4.1.1}
\int_{-\infty}^{x}\left[\psi_{j}^{(0)}(y)\right]^2dy=\frac{I_j(x)}{I_j(x\rightarrow \infty)},
\end{equation}
where
\begin{equation} \label{4.1.2}
I_j(x)=\sum\limits_{l,m=0}^{\lfloor\frac{j}{2}\rfloor}\frac{\left(-1\right)^{m+l}2^{2(j-m-l)-1}}{m!\ l!\ (j-2m)!\ (j-2l)!}
\begin{cases}
\Gamma\left(j-m-l+\frac{1}{2}\right)+\gamma\left(j-m-l+\frac{1}{2},\zeta^2\right),\ \zeta\geq 0,
\\
\Gamma\left(j-m-l+\frac{1}{2}\right)-\gamma\left(j-m-l+\frac{1}{2},\zeta^2\right),\ \zeta<0,
\end{cases}
\end{equation}
with $\lfloor x\rfloor$ being the floor function, $\Gamma(s)$ the gamma function, $\gamma(s,x)$ the lower incomplete gamma function, and $\zeta=\sqrt{\omega/2}\left(x+2\kappa/\omega\right)$. Note that $I_j(x\rightarrow \infty)$ turns out to be
\begin{equation}
\label{4.1.3}
I_j(x\rightarrow \infty)=\sum\limits_{l,m=0}^{\lfloor\frac{j}{2}\rfloor}\frac{\left(-1\right)^{m+l}2^{2(j-m-l)}}{m!\ l!\ (j-2m)!\ (j-2l)!}\Gamma\left(j-m-l+\frac{1}{2}\right).
\end{equation}

\begin{figure}[ht]
\centering
\subfigure[]{\includegraphics[width=8.2cm, height=5.5cm]{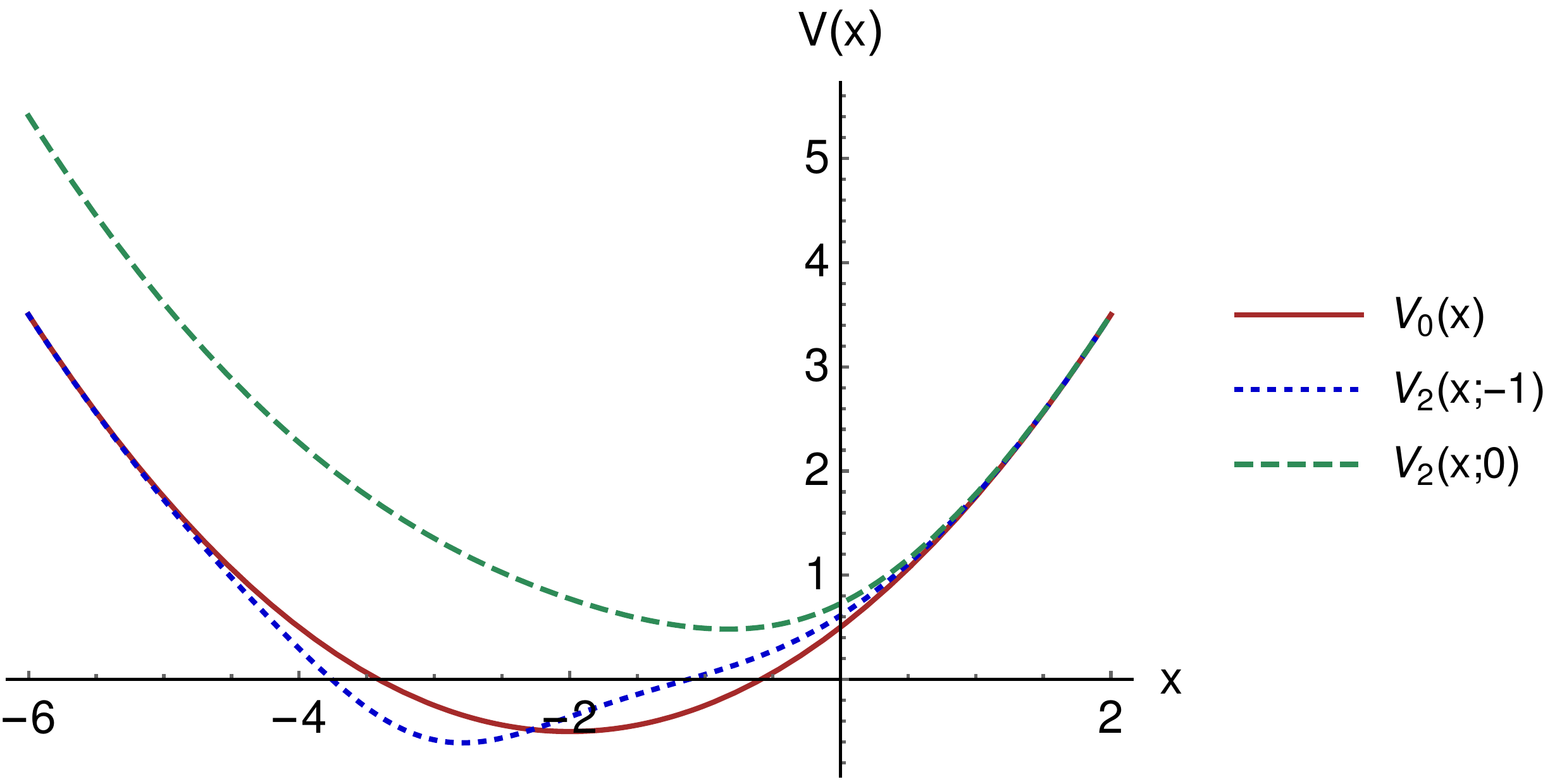}}
\subfigure[]{\includegraphics[width=8.2cm, height=5.5cm]{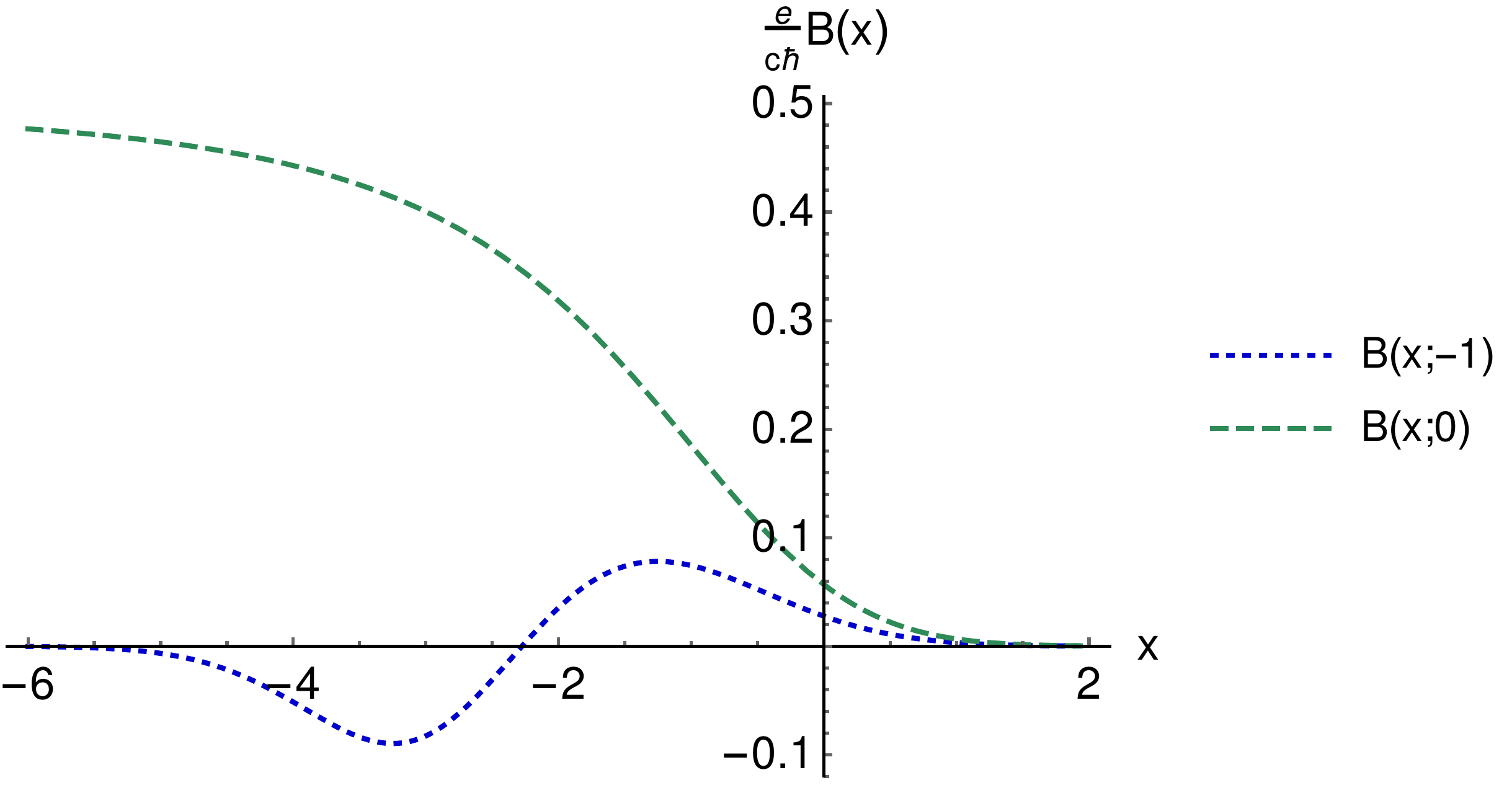}}
\caption{(a) Plot of the potentials $V_2(x;-1)$ (isospectral case), $V_2(x;0)$ (limit case) and $V_0(x)$ (continuous line) for the confluent algorithm applied to the shifted harmonic oscillator, as well as the associated magnetic fields (b). The parameters were taken as $\omega=\kappa=1$.} \label{F-4-1-1}
\end{figure}

By using now equations~(\ref{4.2},\ref{4.1.1}-\ref{4.1.3}) it is straightforward to obtain the function $\eta(x)$ and then the SUSY partner potential $V_2(x;\mathrm{w}_0)$. 
In particular, for $j=0$ the SUSY partner potential $V_2(x;\mathrm{w}_0)$ is given by equation~\eqref{2.16}, where the corresponding one-parametric family of associated magnetic fields turns out to be
\begin{equation}
 \label{4.1.4}
\mathcal{B}(x;\mathrm{w}_0)=\frac{c\hbar\omega}{e}\left\{\frac{e^{-2\zeta^2}}{\pi  \left[\text{erfc}\left(\zeta\right)-2+2\mathrm{w}_0\right]^2}-\frac{\zeta e^{-\zeta^2}}{\sqrt{\pi} \left[\text{erfc}\left(\zeta\right)-2+2\mathrm{w}_0\right]}\right\},
\end{equation}
where $\text{erfc}(x)$ is the complementary error function. Plots of $V_0(x)$, its SUSY partner potentials $V_2(x;-1)$ and $V_2(x;0)$ as well as the associated magnetic fields are shown in Figure \ref{F-4-1-1}.

\begin{figure}[ht]
\centering
\subfigure[]{\includegraphics[width=8.2cm, height=5.5cm]{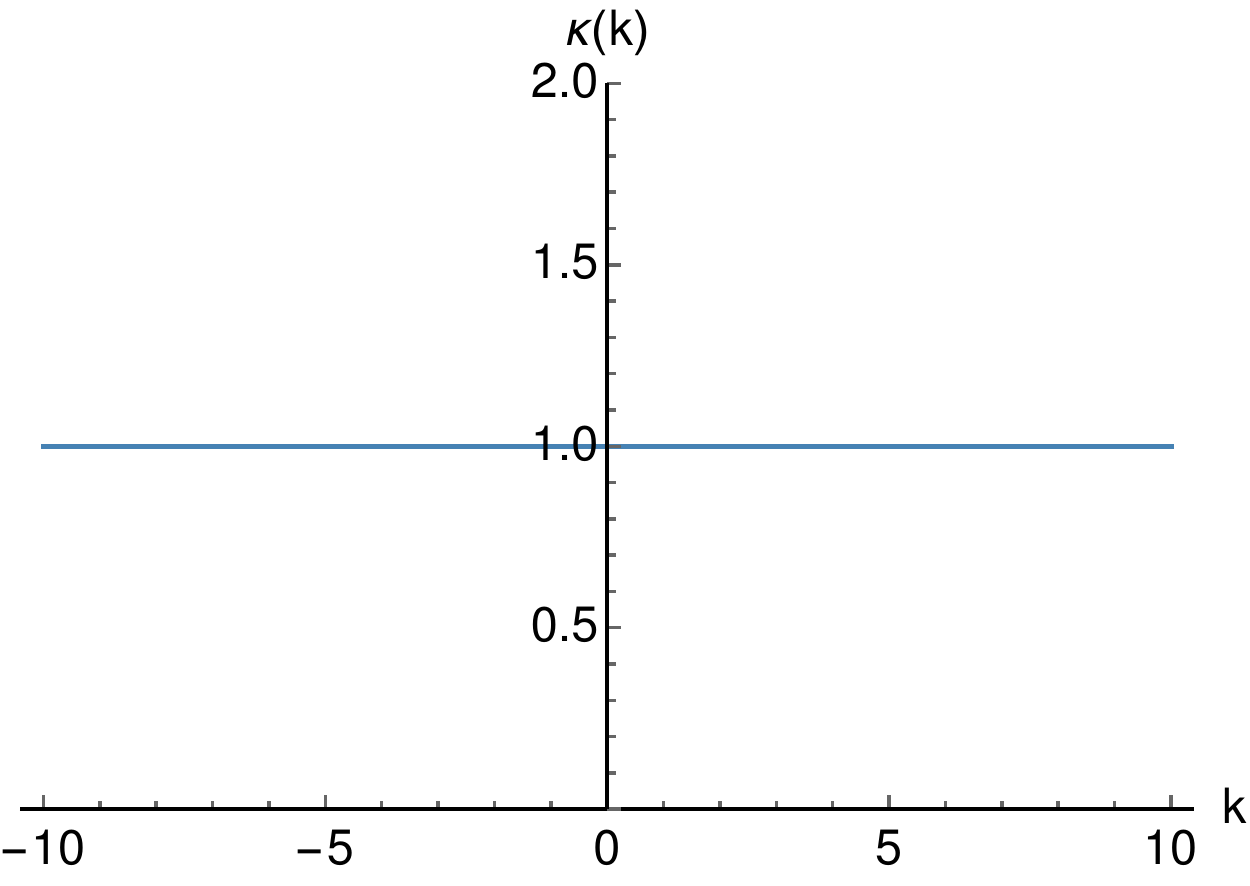}}
\subfigure[]{\includegraphics[width=8.2cm, height=5.5cm]{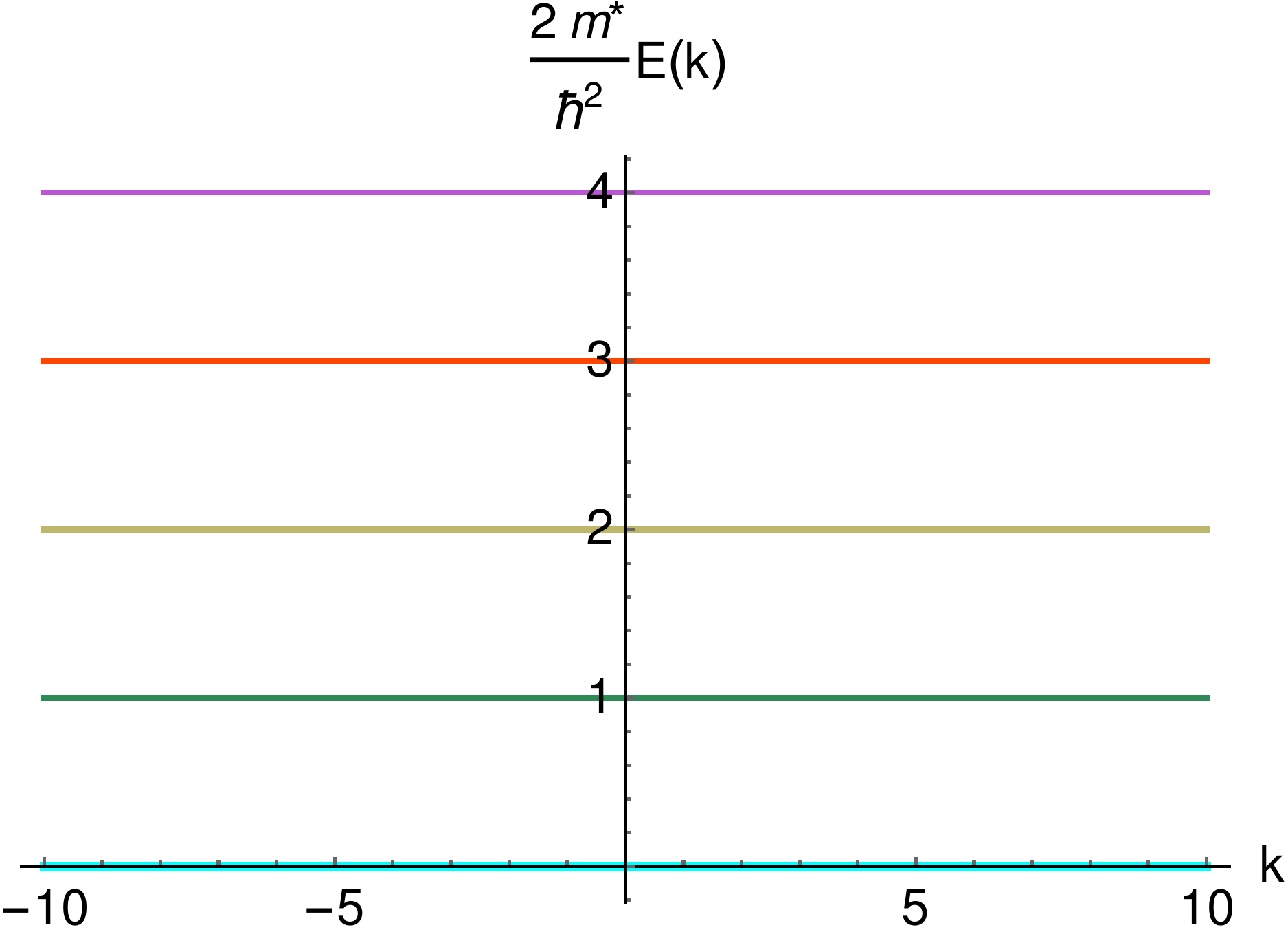}}
\caption{Parameter $\kappa$ (a) and electron energies $E_n$ (b) as functions of the wavenumber $k$ for the confluent algorithm applied to the shifted harmonic oscillator with $\omega=1$.} \label{F-4-1-2}
\end{figure}

\begin{figure}[ht]
\centering
\subfigure[]{\includegraphics[width=8.2cm, height=5.5cm]{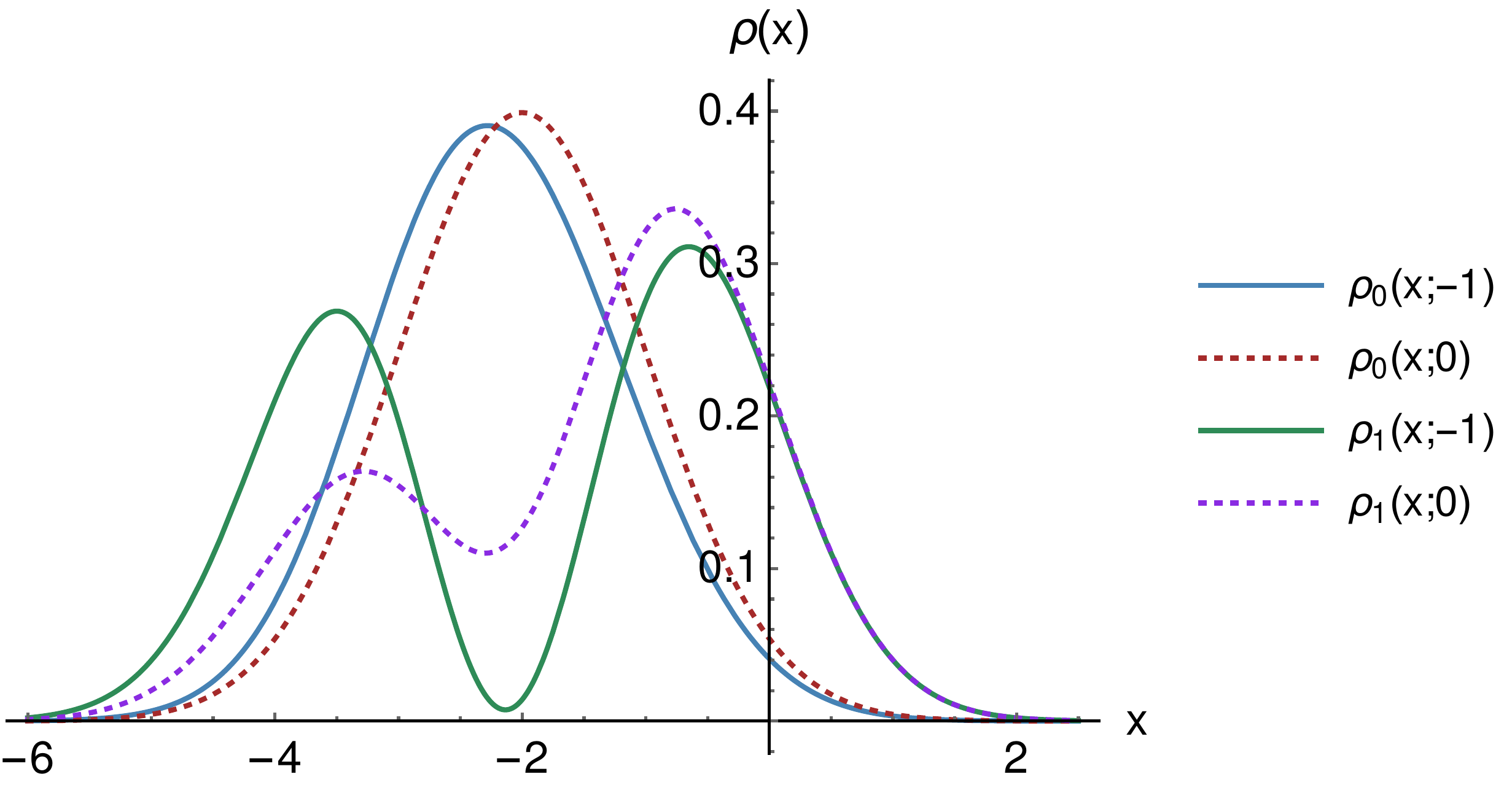}}
\subfigure[]{\includegraphics[width=8.2cm, height=5.5cm]{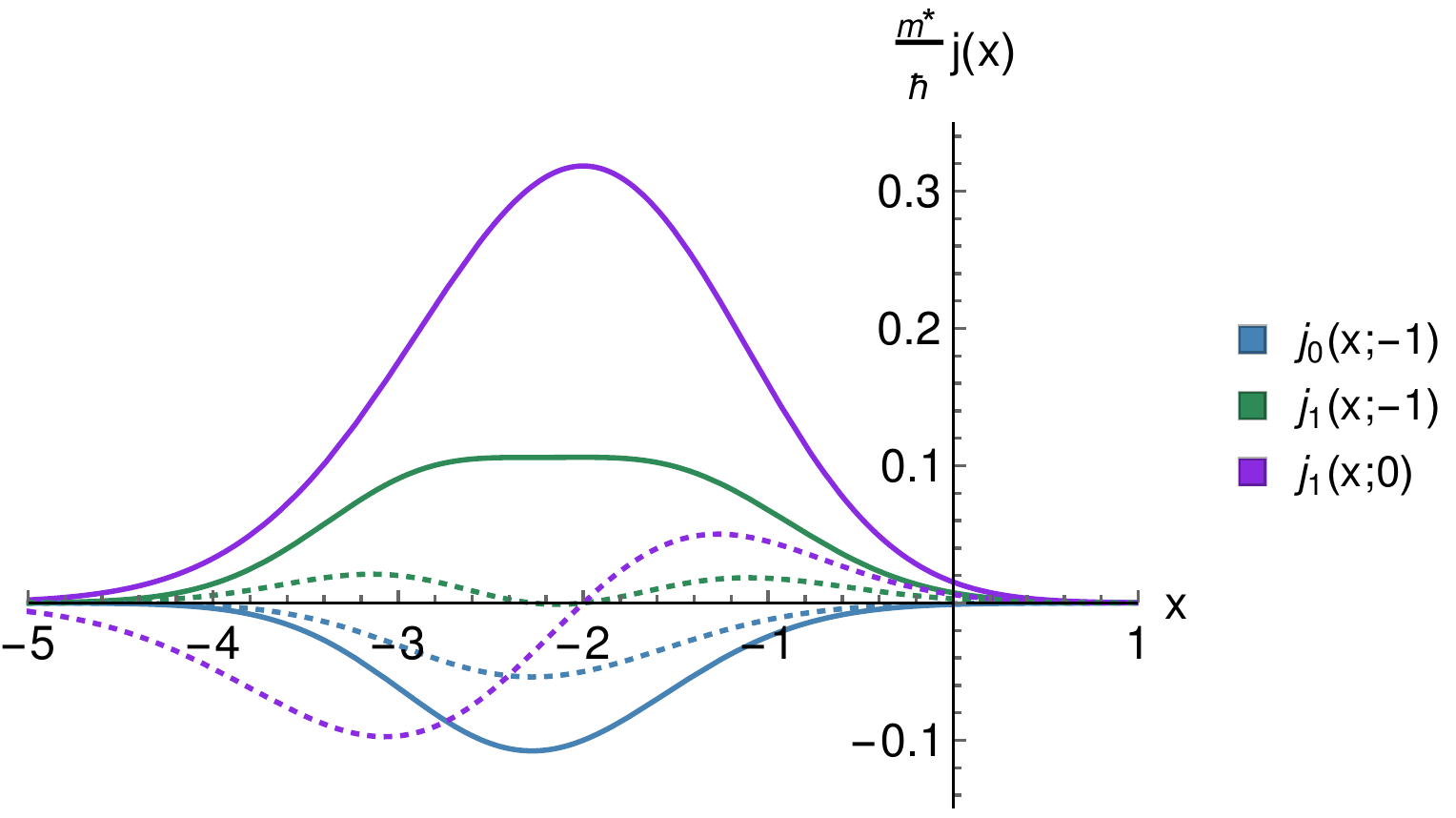}}
\caption{Probability densities in both, isospectral and limit cases (a), as well as current densities (b) for the confluent algorithm applied to the shifted harmonic oscillator. The parameters were taken as $\omega=\kappa=1$ and $k=0$. In (b) the current density $J_{x}$ corresponds to the dotted lines while $J_{y}$ does for the continuous lines.} \label{F-4-1-3}
\end{figure}

Going back to our original problem, the energy eigenvalues for electrons in bilayer graphene in the magnetic field (\ref{4.1.4}) are given by  
\begin{equation} \label{4.1.5}
E_n=\frac{\hbar^2\omega}{2m^{*}}n,\qquad n=0,1,\dots
\end{equation}
We can see that these eigenvalues are equidistant, which is similar to what happens for the energies \eqref{3.1.2} of the auxiliary problem. Moreover, since $j=0$ all of them are non-degenerate. However, if we would take $j\geq 1$, despite the spectrum would remain equidistant the first $j$ excited state energies would be doubly degenerate. Note also that $E_n$ does not depend on the parameter $\kappa$ nor of the wavenumber $k$, since now both quantities are not related (check Figure~\ref{F-4-1-2}). Due to this fact, the potentials and magnetic fields shown in Figure \ref{F-4-1-1}~ are independent of the wavenumber $k$. The probability densities are as well $k$-independent (see Figure~\ref{F-4-1-3}), but the probability currents indeed depend on $k$ (check equation \eqref{2.18}).

\subsection{Trigonometric Rosen-Morse potential}
Let us analize next the trigonometric Rosen-Morse potential of equation~\eqref{3.2.1}, for which the eigenfunctions and eigenvalues are given in equations \eqref{3.2.2} and \eqref{3.2.3}, respectively. As in the previous example, we take as seed solution the bound state $\psi_{j}^{(0)}(x)$, then $\mathrm{w}(x)$ can be written once again as in equation \eqref{4.1.1}, but now $I_j(x)$ looks like
\begin{equation}
\label{4.2.1}
\begin{aligned}
I_j(x)=&\sum\limits_{l,m=0}^{j}(-1)^{m+l}\binom{j}{m}\binom{j}{l}\frac{(1+q-p+j)_l(1+q-p+j)_m}{\Gamma(l+1-p)\Gamma(m+1-p)}\bigg[e^{-i\frac{p+q}{2}\left[\pi-2\theta(x)\right]}\ e^{-i(p-q)\theta(x)}
\\
&\times\ _{2}F_{1}\left(q,m+l+q-p;1+q;-e^{i2\theta(x)}\right)-\frac{\Gamma(1+q)\Gamma(p-q-m-l+1)}{\Gamma(p+1-m-l)}e^{-i\frac{p-q}{2}\pi}\bigg],
\end{aligned}
\end{equation}
where $p=s+j+ia$, $q=-s-j+ia$, $\theta(x)=\text{arctan}\left(\cot (\alpha x)\right)$, $_{2}F_{1}\left(a,b;c;x\right)$ is the hypergeometric function and $I_j(x\rightarrow\pi)$ turns out to be  
\begin{equation}
\label{4.2.2}
\begin{aligned}
I_j(x\rightarrow\pi)&=\sum\limits_{l,m=0}^j (-1)^{l+m} \binom{j}{l} \binom{j}{m} \frac{(1+q-p+j)_l(1+q-p+j)_m}{\Gamma (l+1-p)\Gamma (m+1-p)}\frac{\Gamma (1+q) \Gamma (p-q+1-l-m)}{\Gamma (p+1-m-l)}
\\
&\times\left(e^{-i(p+q)\pi} e^{i\frac{p-q}{2}\pi}-e^{-i\frac{p-q}{2}\pi)}\right).
\end{aligned}
\end{equation}

\begin{figure}[ht]
\centering
\includegraphics[width=8.2cm, height=5.5cm]{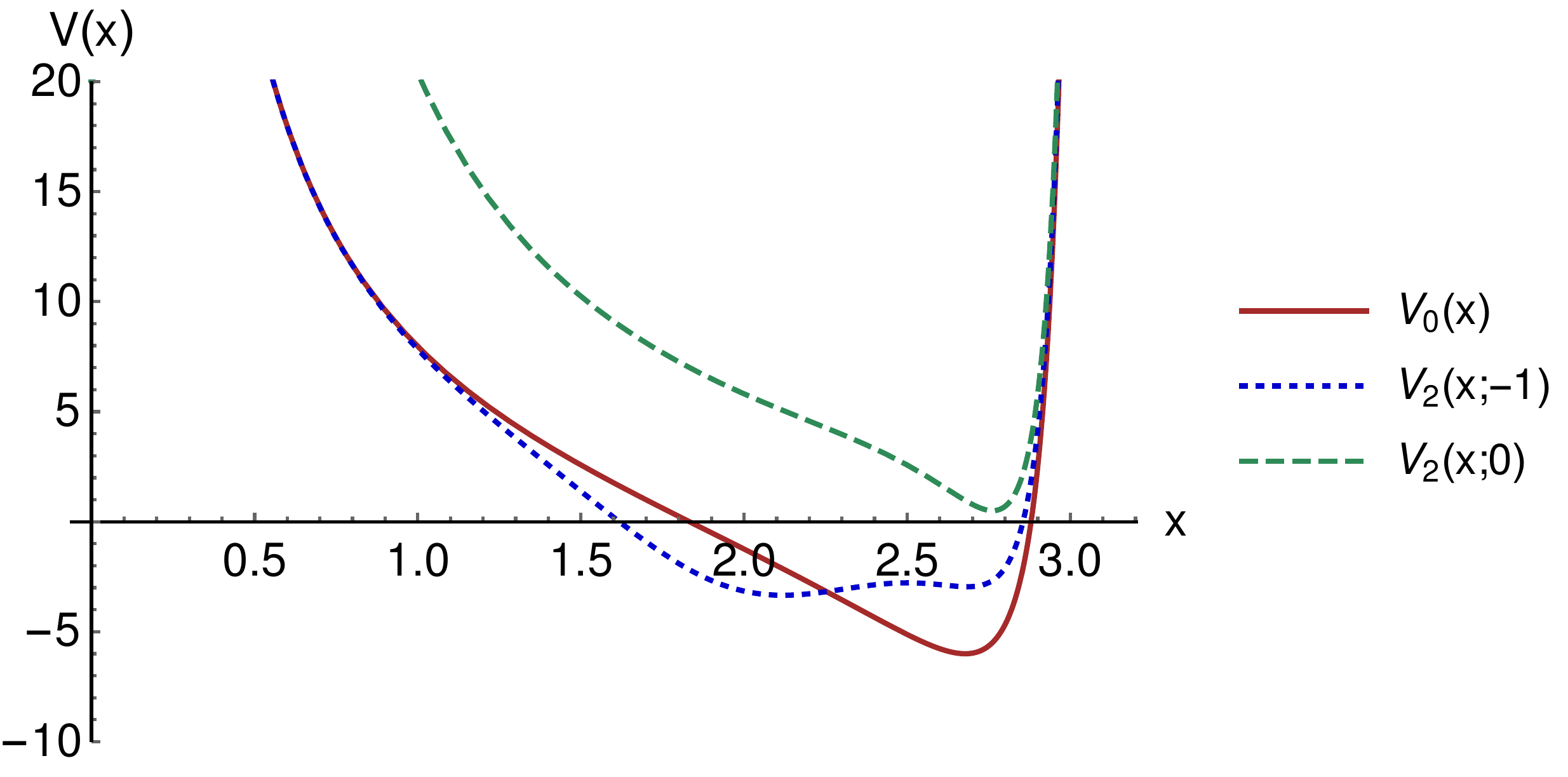}
\includegraphics[width=8.2cm, height=5.5cm]{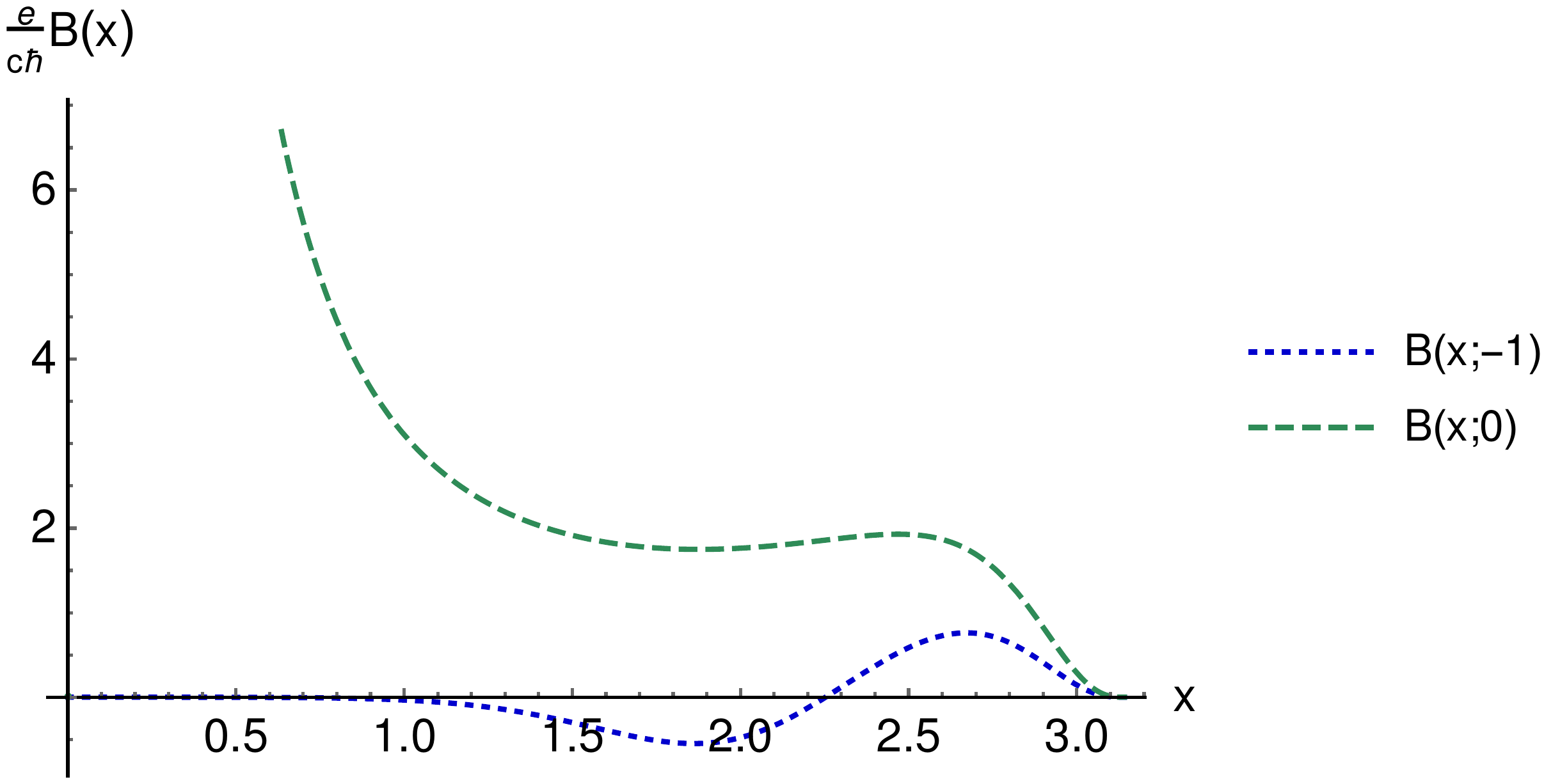}
\caption{(a) Plot of the potentials $V_2(x;-1)$ (isospectral case), $V_2(x;0)$ (limit case) and $V_0(x)$ (continuous line) as well as the associated magnetic fields (b) for the confluent algorithm applied to the trigonometric Rosen-Morse potential. The parameters were taken as $D=2$, $\kappa=-2$ and $\alpha=1$}\label{F-4-2-1}
\end{figure}

By taking now $j=0$, the SUSY partner potential $V_2(x;\mathrm{w}_0)$ is given by equation~\eqref{2.16}, where the corresponding one-parametric family of magnetic fields becomes
\begin{equation}
\label{4.2.3}
\begin{aligned}
&\mathcal{B}(x;\mathrm{w}_0)=-2q\frac{c\hbar}{e}\alpha^2\Gamma (1+p) \left(1+e^{i2\theta(x)}\right)^{p-q} \bigg\{\Big[\Gamma(1+q) \Gamma (p-q+1) \left[1-i2\mathrm{w}_0\sin (q\pi)\right] e^{-i2q\theta(x)}
\\
&-\Gamma\left(1+p\right)\, _2F_1\left(q,q-p;1+q;e^{-i2\alpha x}\right)\Big]^{-2}\Big[q \Gamma (1+p)\Big(\left(1-e^{-i2\alpha  x}\right)^{p-q}-\, _2F_1\left(q,q-p;1+q;e^{-i2\alpha x}\right)\Big) 
\\
&-\frac{p-q}{1+e^{-i2\theta(x)}}\Big(\Gamma (1+p) \, _2F_1\left(q,q-p;1+q;e^{-i2\alpha x}\right)-\Gamma (p-q+1) \Gamma (1+q) \left[1-i2\mathrm{w}_0\sin (q\pi)\right] e^{-i2q\theta(x)}\Big)
\\
&+q\Gamma (p-q+1) \Gamma (1+q) \left[1-i2\mathrm{w}_0\sin (q\pi)\right] e^{-i2q\theta(x)}\bigg]\bigg\}.
\end{aligned}
\end{equation}
Note that now these potentials and magnetic fields depend on the wavenumber $k$, as it is shown in Figure \ref{F-4-2-1}. 

\begin{figure}[ht]
\centering
\includegraphics[width=8.2cm, height=5.5cm]{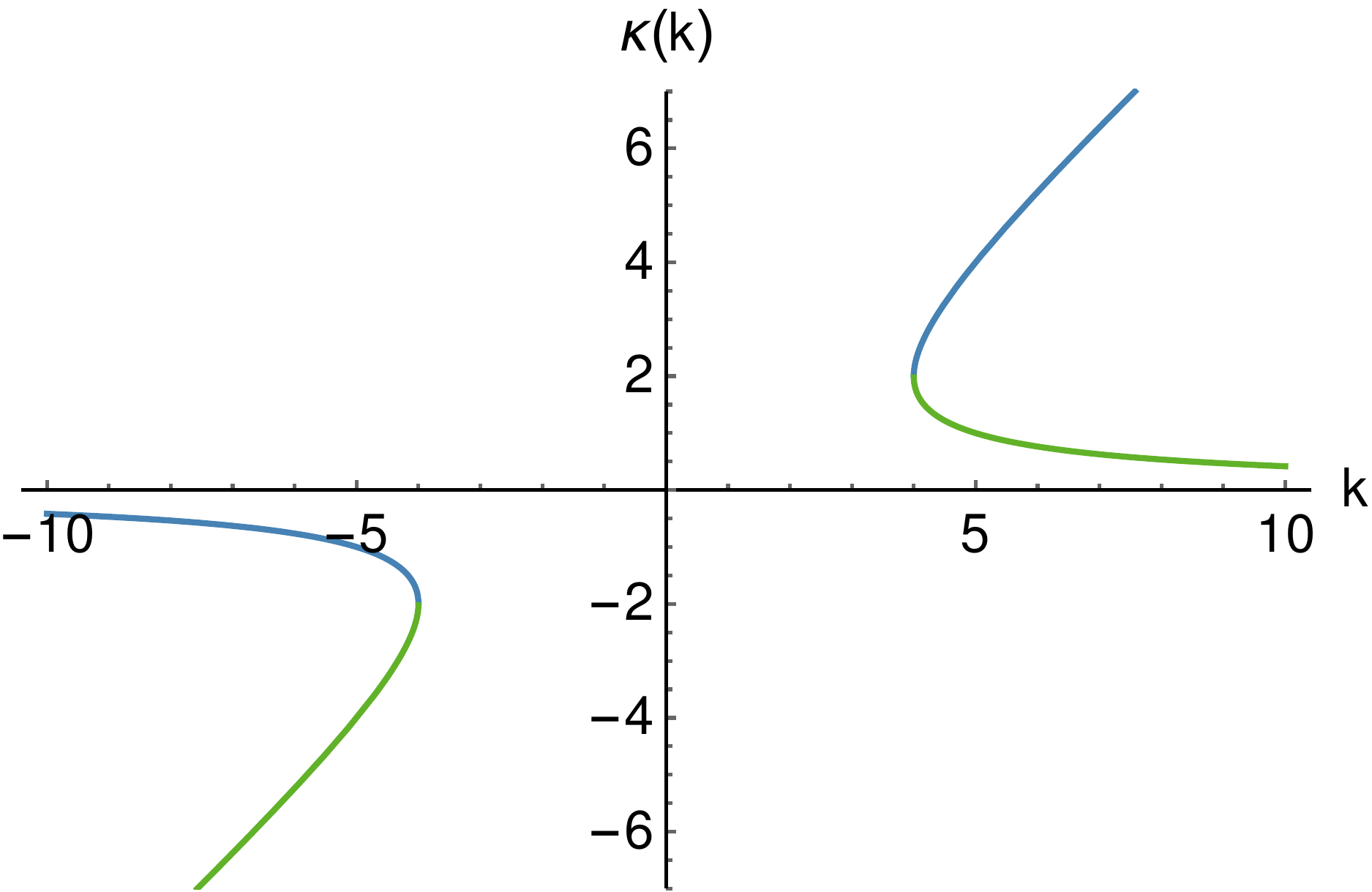}
\includegraphics[width=8.2cm, height=5.5cm]{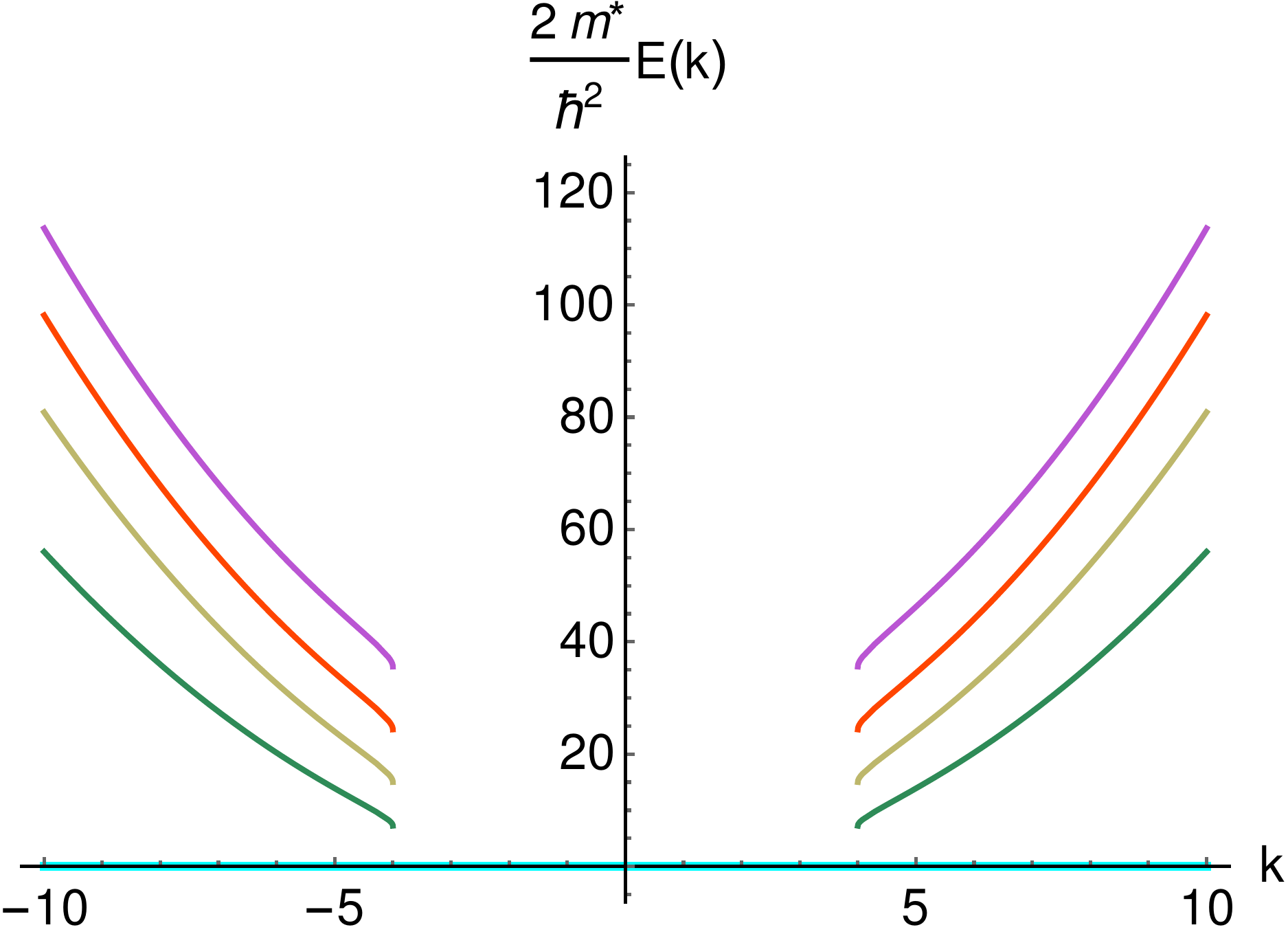}
\caption{The two possible solutions of the parameter $\kappa(k)$ (a) and electron energies as functions of the wavenumber $k$ (b) for the confluent algorithm applied to the trigonometric Rosen-Morse potential. The parameter were taken as $D=2,\ \alpha=1$.}\label{F-4-2-2}
\end{figure}

For these magnetic fields the energy eigenvalues for electrons in bilayer graphene are given by:
\begin{equation} \label{4.2.4}
E_n=\frac{\hbar^2}{2m^{*}}\left[\kappa^2-D^2+\left(D+n\alpha\right)^2-\frac{\kappa^2D^2}{\left(D+n\alpha\right)^2}\right].
\end{equation}
Once again, up to a factor these eigenvalues are the same as the eigenenergies for the auxiliary Hamiltonian of equation~\eqref{3.2.3}. Note also that they are non-degenerate (see Figure \ref{F-4-2-2}). However, in contrast to what happens for the harmonic oscillator potential, in this case there is a dependence of $E_n$ on the wavenumber $k$, induced by the fact that $\kappa$ and $k$ are now linked as follows
\begin{equation} \label{4.2.5}
k=\kappa +\frac{4}{\kappa}.
\end{equation}

\begin{figure}[ht]
\centering
\includegraphics[width=8.2cm, height=5.5cm]{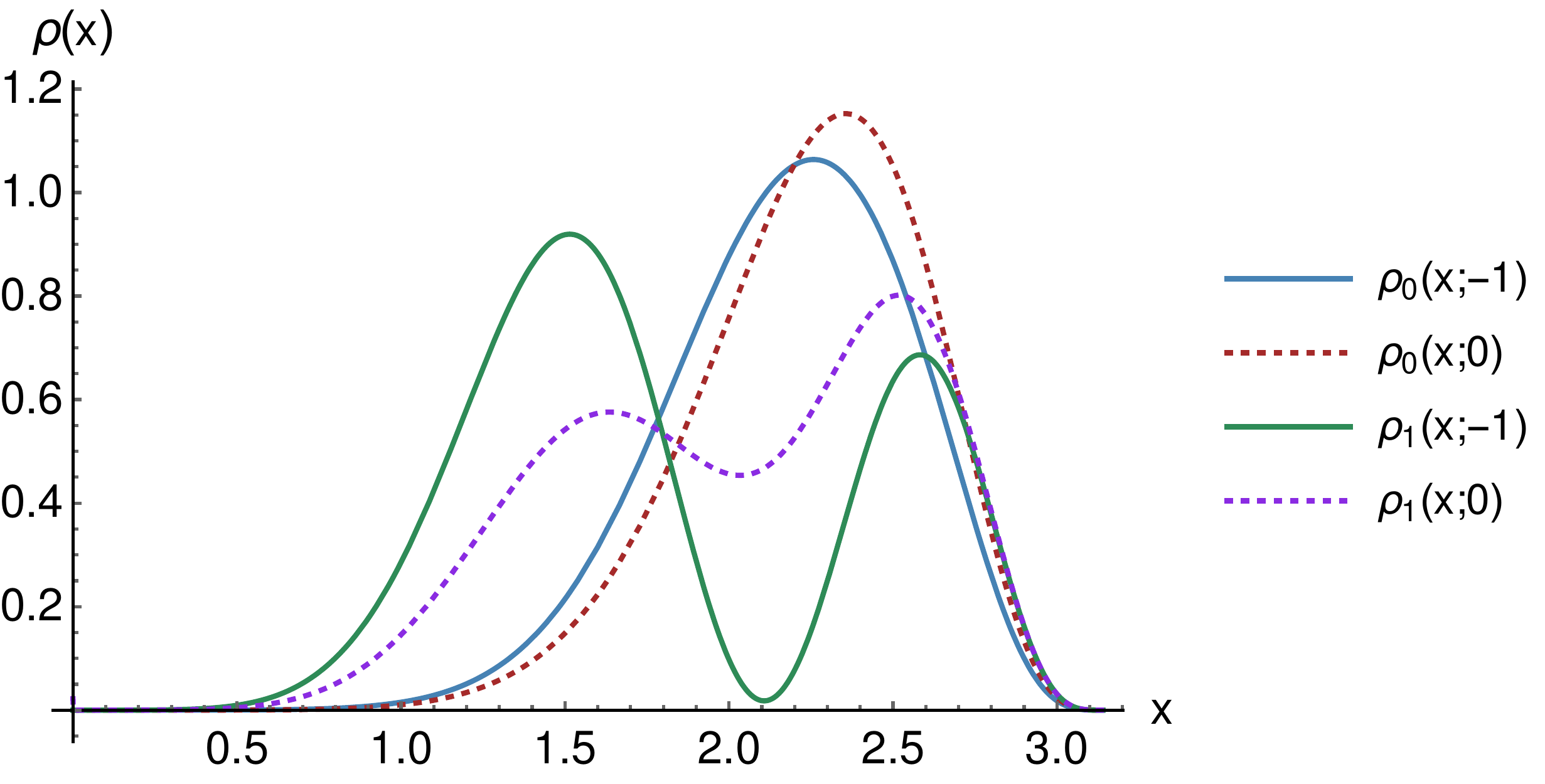}
\includegraphics[width=8.2cm, height=5.5cm]{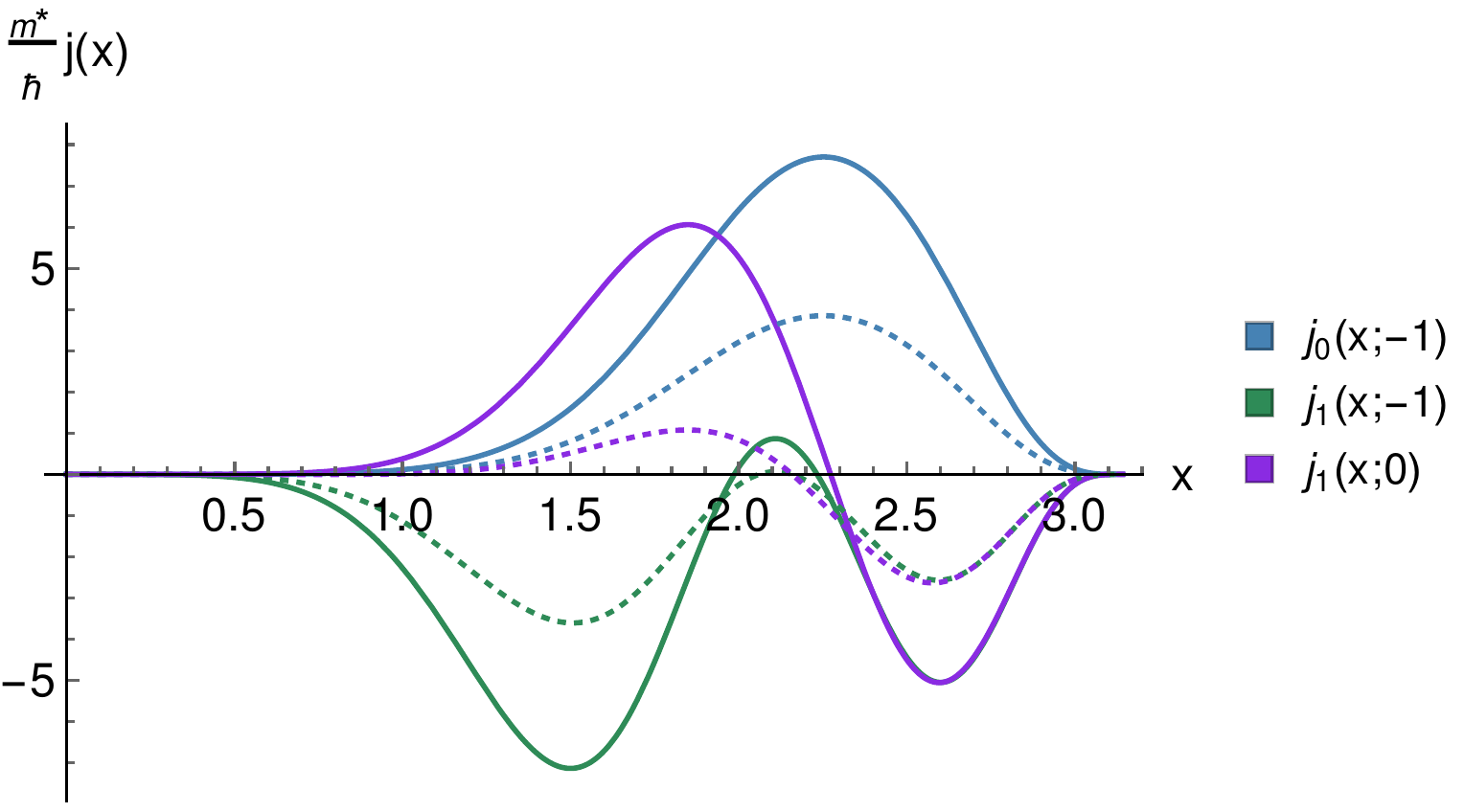}
\caption{Probability (a) and current densities (b) in both, isospectral and limit cases, for the confluent algorithm applied to the trigonometric Rosen-Morse potential. The parameter were taken as $D=2,\ \alpha=1$ and $\kappa=-2$. In (b) the current density $J_{x}$ corresponds to the dotted lines while $J_{y}$ does for the continuous lines.}\label{F-4-2-3}
\end{figure}

It is important to stress that this equation is valid only for $D=2$ and $\alpha=1$, since the calculation for arbitrary values of these parameters is 
difficult. By solving now equation~\eqref{4.2.5} for $\kappa(k)$ we will get two different solutions (see Figure \ref{F-4-2-2}). Even though both are mathematically correct, physically the electron energies \eqref{4.2.4} must depend on each of these solutions in a way that the quadratic behavior with respect to the momentum remains valid. Consequently, only in the domain $(-4,4)$ the ground state exists. The probability and current densities for some eigenfunctions are plotted in Figure \ref{F-4-2-3}, which depend on the parameter $\kappa$ and thus they are non-trivial functions of $k$. 


\subsection{Hyperbolic Rosen-Morse potential}
Finally, let us take the hyperbolic Rosen-Morse potential (also called Rosen-Morse II potential) as the $V_0(x)$ of equation (\ref{3.3.1}). If we choose as seed solution the $j$-th eigenstate of equation (\ref{3.3.3}), the integral appearing in $\mathrm{w}(x)$ turns out to be
\begin{equation}
I_j(x)=\frac{G(\zeta,p,q,j)+G(1,q,p,j)}{G(1,p,q,j)+G(1,q,p,j)},
\label{4.3.1}
\end{equation}
where $\zeta=\text{tanh}(\alpha x)$, $s=\frac{D}{\alpha}$, $a_j=\frac{\kappa}{\alpha(D-\alpha j)}$, $p=s-j+a_j$, $q=s-j-a_j$ and
\begin{equation}
\begin{aligned}
&G(a,b,c,j)=\Gamma^2(b+j+1)\sum _{l,r=0}^j\left(-\frac{1}{2}\right)^{l+r}\binom{j}{l}\binom{j}{r}\left(\frac{\Gamma(b+c+j+l+1)\Gamma(b+c+j+r+1)}{2(b+l+r)\Gamma(b+l+1)\Gamma(b+r+1)}\right)\Bigg\{2^{b+c+l+r} \\
&\times(b+c+l+r)\left[\text{B}\left(\frac{a+1}{2};c,b+l+r+1\right)-\text{B}\left(\frac{1}{2};c,b+l+r+1\right)\right]-(1-a)^{b+l+r}(1+a)^c+1\Bigg\},
\end{aligned}
\label{4.3.2}
\end{equation}
with $\text{B}(x;a,b)$ being the incomplete beta function. As an example, by taking now $j=0$ we obtain the one-parametric family of SUSY partner potentials $V_2(x;\mathrm{w}_0)$ of equation~\eqref{2.16} with the following magnetic field expression: 
\begin{equation}
\begin{aligned}
&\mathcal{B}(x;\mathrm{w}_0)=\frac{c\hbar}{e}\alpha^{2}pq(1-\zeta)^{p} (1+\zeta)^{q}\Bigg\{q(1-\zeta)^p(1+\zeta)^q+(\mathrm{w}_0-1) (p+q)-2^{p+q} \bigg[p q (1-2 \mathrm{w}_0) \text{B}(q,p)
\\
&+(p+q)\bigg(q \text{B}\left(\frac{\zeta+1}{2};q,p+1\right)+(\mathrm{w}_0-1) \left(p \text{B}\left(\frac{1}{2};p,q+1\right)+q \text{B}\left(\frac{1}{2};q,p+1\right)\right)\bigg)\bigg]\Bigg\}^{-2}\Bigg\{(p+q)
\\
&\times\Big(q(1-\zeta)^{1+p}(1+\zeta)^{q}-(p-q+(p+q)\zeta)(-1+\mathrm{w}_{0})\Big)+2^{p+q}\big(p-q+(p+q)\zeta\big)\bigg[pq(1-2\mathrm{w}_{0})\text{B}(q,p)
\\
&+(p+q)\left((-1+\mathrm{w}_{0})\left(p\text{B}\left(\frac{1}{2};p,1+q\right)+q\text{B}\left(\frac{1}{2};q,1+p\right)\right)+q\text{B}\left(\frac{1+\zeta}{2};q,1+p\right)\right)\bigg]\Bigg\}.
\end{aligned}
\end{equation}

The eigenenergies for the electron in bilayer graphene under these magnetic fields become 
\begin{equation}
E_n=\frac{\hbar^2}{2m^*}\left[{D}^{2}+\kappa^{2}-\left({D}-n\alpha\right)^{2}-\frac{\kappa^{2}{D}^{2}}{\left({D}-n\alpha\right)^{2}}\right], \quad n=0,1,\dots,N.
\label{4.3.5}
\end{equation}
Nevertheless, a big difference in the associated eigenfunctions appears, as we have discussed previously.

\begin{figure}[ht]
\centering
\includegraphics[width=8.2cm, height=5.5cm]{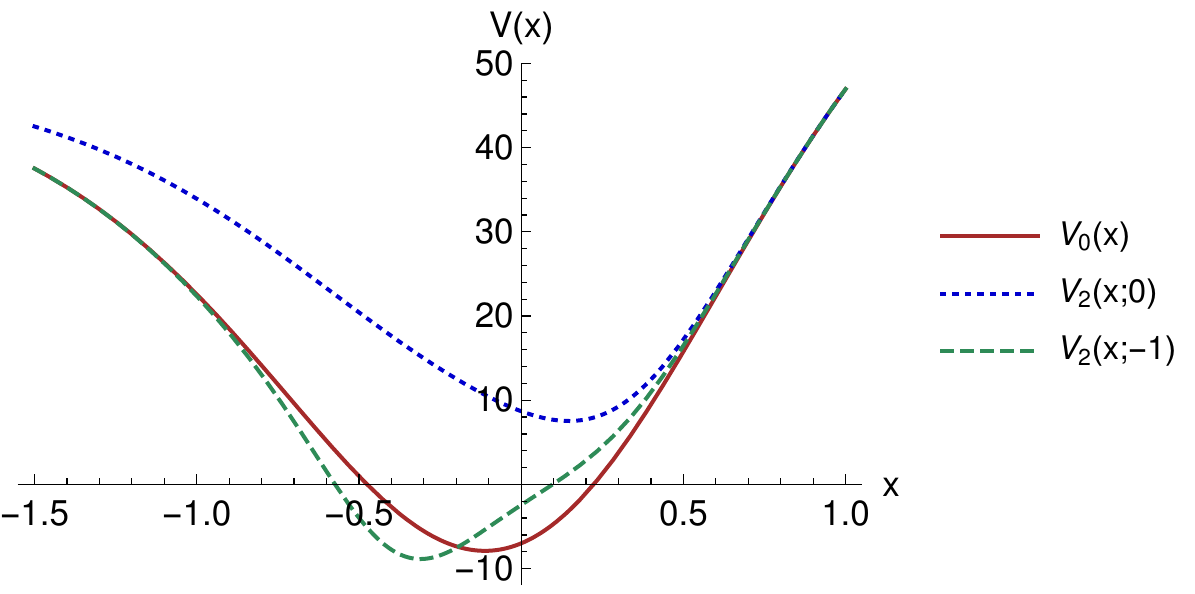}
\includegraphics[width=8.2cm, height=5.5cm]{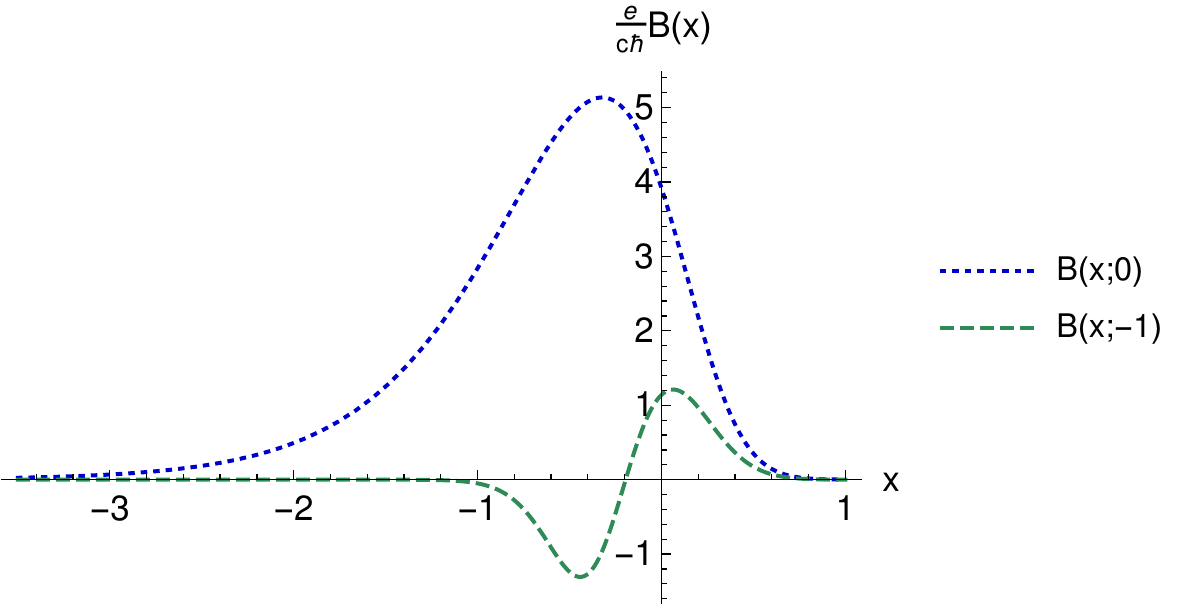}
\caption{Plot of the hyperbolic Rosen-Morse potential $V_0$ and its confluent second-order SUSY partner potentials $V_2$ in the isospectral case ($w_0=-1$) and in the limit case ($w_0=0$) (left), as well as the associated magnetic fields (right). The parameters were taken as $D=8$, $\kappa=1$ and $\alpha=1$} \label{F-4-3-1}
\end{figure}

\begin{figure}[ht]
\centering
\includegraphics[width=8.2cm, height=5.5cm]{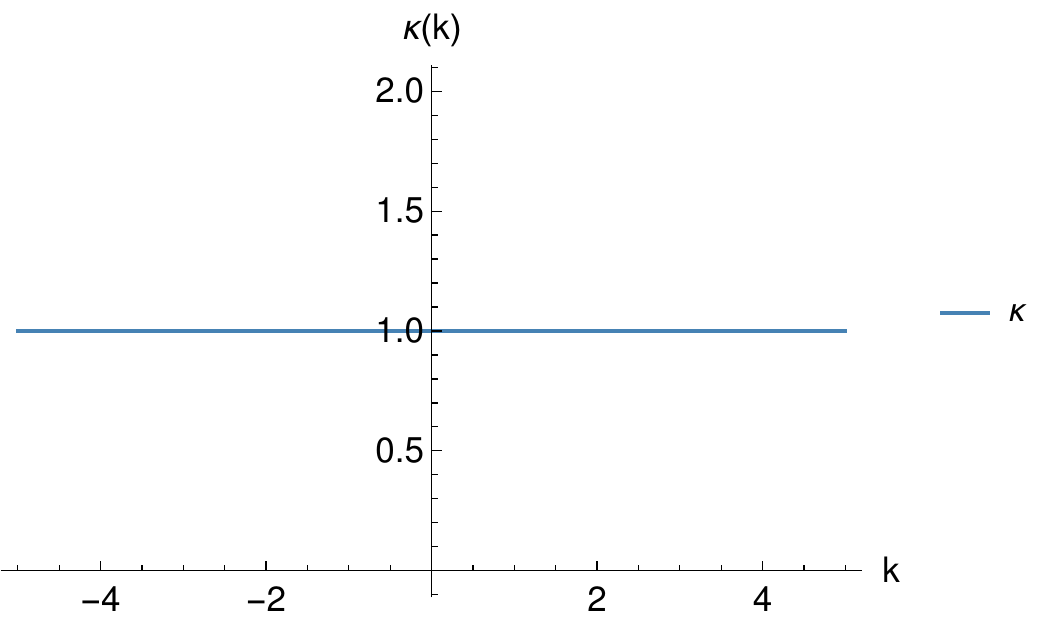}
\includegraphics[width=8.2cm, height=5.5cm]{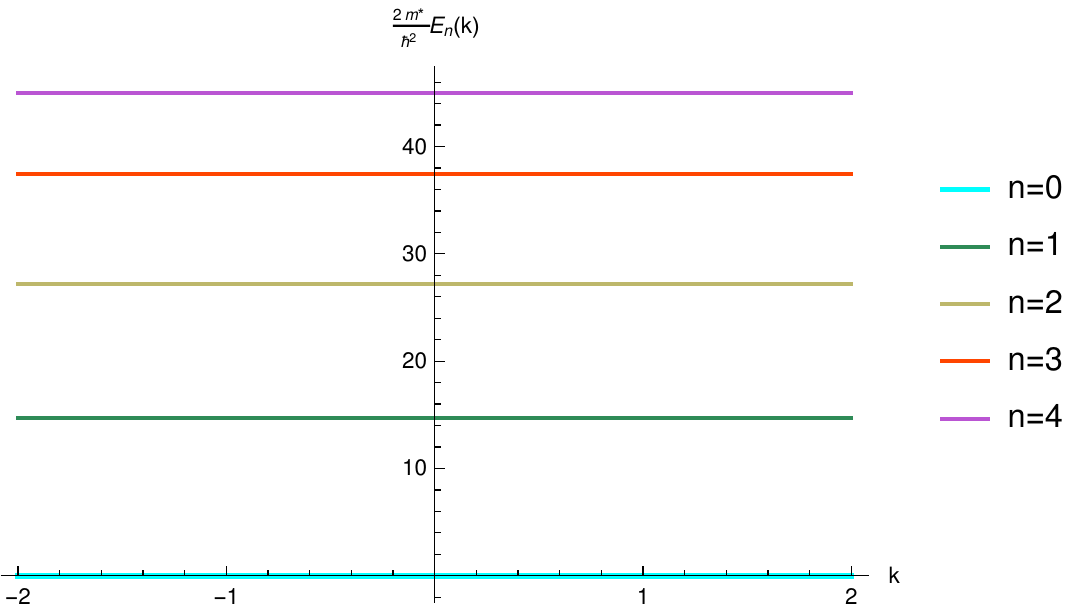}
\caption{Plot of $\kappa$ as function of the wavenumber $k$ (left), and some electron energies $E_n$ (right) for the confluent algorithm applied to the hyperbolic Rosen-Morse potential. The parameters were taken as $D=8$, $\alpha=1$ and $\kappa=1$} \label{F-4-3-2}
\end{figure}

\begin{figure}[ht]
\centering
\includegraphics[width=8.2cm, height=5.5cm]{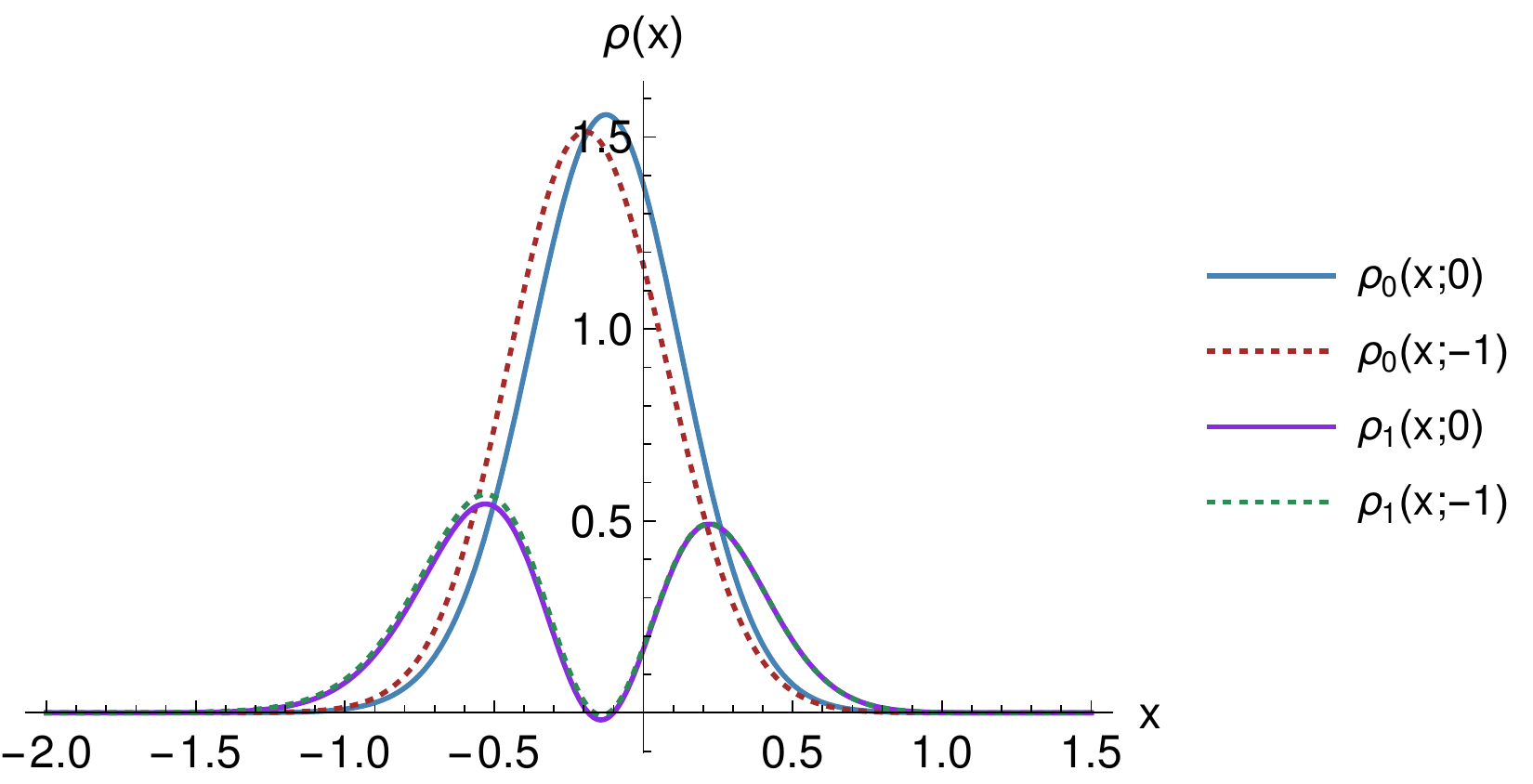}
\includegraphics[width=8.2cm, height=5.5cm]{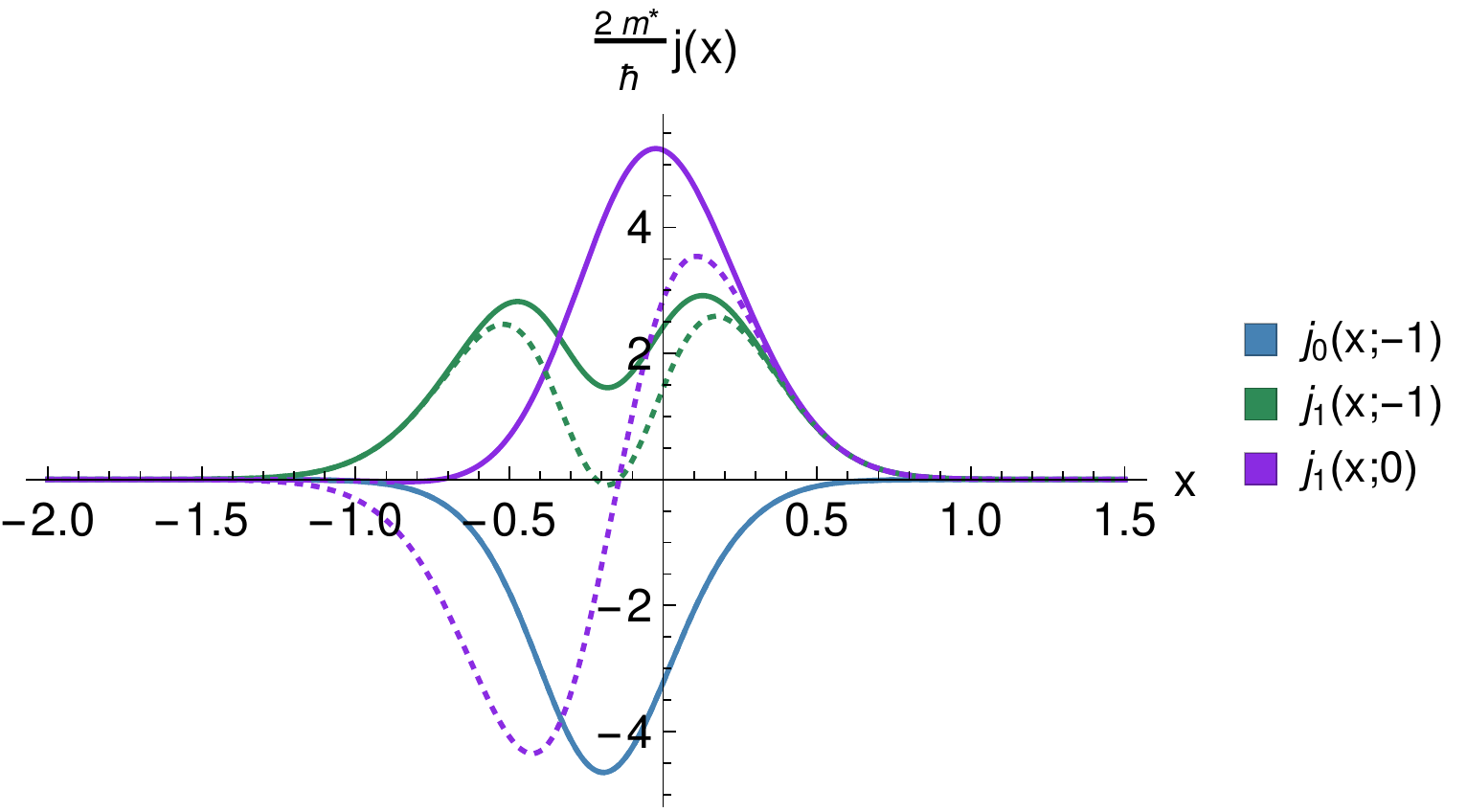}
\caption{Probability density (a) and current densities $J_{x}$ (dotted lines) and $J_{y}$ (continuous lines)(b) in both, isospectral and limit cases, for the confluent algorithm applied to the hyperbolic Rosen-Morse potential. The parameters were taken as $D=8$, $\alpha=1$, $\kappa=1$ and $k=1$.} \label{F-4-3-3}
\end{figure}

Once again, these eigenvalues are proportional to the auxiliary eigenenergies of equation~(\ref{3.3.3}), all of them being non-degenerate, and the index $n$ supplies the standard ordering for this set. Besides, $\kappa$ and $k$ are not related for $\mathrm{w}_0\neq0$, but for $\mathrm{w}_0=0$ a non-trivial dependence appears. In Figure (\ref{F-4-3-1}) we can see plots of the SUSY partner potentials $V_2(x;\mathrm{w}_0)$ and the associated magnetic fields. The electron energies do not depend on $k$ (see Figure~\ref{F-4-3-2}), but they are still functions on $\kappa$, a behavior which is similar to the first case of this section. Finally, the probability and current densities are drawn in Figure \ref{F-4-3-3}, the former not being functions of $k$ any longer but the current density is.

\section{Conclusions}

We have shown that the second-order supersymmetric quantum mechanics allows to address in a natural way the problem of eigenvalues for the effective Hamiltonian of electrons in bilayer graphene with applied external magnetic fields. Analytic solutions for the Hamiltonian bound states have been obtained through this procedure, and a lot of exotic magnetic fields have been generated during such a process.

We have generated the external magnetic fields of this paper departing from a given exactly solvable potential $V_0$, and its non shape-invariant SUSY partner potentials $V_2$. Two important cases have been addressed in this paper: (i) in the first case the factorization energies were chosen as two consecutive energy levels $\mathcal{E}^{(0)}_j$, $\mathcal{E}^{(0)}_{j+1}$, $j\geq 1$ of the initial potential. Unlike the shape invariant case, which appears as the limit for $j=0$, for $j\geq 1$ there is always at least one two-fold degenerate eigenvalue (the ground state energy). Sometimes it is possible to find more degenerate eigenvalues, but this will depend, in general, on the choice of the parameters of $V_0$. In particular, for the shifted harmonic oscillator there appear $j$ additional two-fold degenerate eigenvalues, independently of this parameters choice. (ii) In the confluent case the parameters choice defines as well the degeneracy of the eigenvalues, but for the oscillator case there will be always $j$ two-fold degenerate eigenvalues. It is worth to note that this degeneracy appears only for the excited state eigenvalues, since in the confluent case the ground state energy is always non-degenerate.

For the case when two consecutive energy levels $\mathcal{E}^{(0)}_j$, $\mathcal{E}^{(0)}_{j+1}$ of $H_0$ are chosen, the energy eigenvalues $E_n$ for $n<j$ show a rate of increase (decrease) with respect of $k$ which is bigger (smaller) than the corresponding behavior for $E_n$ with $n>j+1$. This is the reason why the position in the spectrum of these electron energies depends on the wavenumber $k$. This does not happen for the displaced harmonic oscillator, since the electron energies in this case are independent of $k$. 

It is important to emphasize that the relationship between the parameter $\kappa$ and the wavenumber $k$ is obtained from analysing the $\eta(x)$ function (see Appendix \ref{apendice}~), a calculation which, in general, is complicated. Finally, we must mention also that, by construction, the wavenumber $k$ is fixed from the very beginning, then the magnetic profile and the eigenfunctions of the Hamiltonian of equation~(\ref{extra2}) in general will depend of $k$. This means that the eigenfunctions and the magnetic profile will change if $k$ is modified. Let us note, however, the exceptional case when the confluent algorithm is applied to the shifted harmonic oscillator, in which the magnetic profile does not depend on the wavenumber so that the electron in bilayer graphene will see always the same magnetic field, independently of the value of $k$.

\bigskip

\noindent{\bf Acknowledgments.} This work was supported by CONACYT (Mexico), project FORDECYT-PRONACES/61533/2020.
JDGM (number 487715) and DOC especially thank to Conacyt for the economic support through the PhD scholarships.

\bibliography{biblio}

\begin{thebibliography}{10}

\bibitem{Katsnelson2007}
M.I. Katsnelson and M.I. Katsnelson.
\newblock {\em Graphene: Carbon in Two Dimensions}.
\newblock Cambridge University Press, 2012.

\bibitem{McCann2013}
Edward McCann and Mikito Koshino.
\newblock The electronic properties of bilayer graphene.
\newblock {\em Reports on Progress in Physics}, 76(5):056503, apr 2013.

\bibitem{Kuru2009}
Ş~Kuru, J.~Negro, and Luis Nieto.
\newblock Exact analytic solutions for a dirac electron moving in graphene
  under magnetic fields.
\newblock {\em Journal of physics. Condensed matter : an Institute of Physics
  journal}, 21:455305, 11 2009.

\bibitem{Milpas2011}
E.~Milpas, M.~Torres, and G.~Murgu\'{i}a.
\newblock Magnetic field barriers in graphene: an analytically solvable model.
\newblock {\em J. Phys.: Condens. Matter}, 23:245304, 2011.

\bibitem{Midya2014}
Bikashkali Midya and David~J Fern{\'{a}}ndez.
\newblock Dirac electron in graphene under supersymmetry generated magnetic
  fields.
\newblock {\em Journal of Physics A: Mathematical and Theoretical},
  47(28):285302, jun 2014.

\bibitem{Erik2017}
E.~D\'{i}az-Bautista and D.J. Fern\'andez.
\newblock Graphene coherent states.
\newblock {\em Eur. Phys. J. Plus}, 132:499, 2017.

\bibitem{Concha2018}
Y.~Concha, A.~Huet, A.~Raya, and D.~Valenzuela.
\newblock Supersymmetric quantum electronic states in graphene under uniaxial
  strain.
\newblock {\em Mat. Res. Express}, 5:065607, 2018.

\bibitem{Roy2018}
D.N. Le, V.H. Le, and P.~Roy.
\newblock Conditional electron confinement in graphene via smooth magnetic
  fields.
\newblock {\em Physica E}, 96:17, 2018.

\bibitem{Erik2019}
E.~D\'{i}az-Bautista, J.~Negro, and L..M. Nieto.
\newblock Partial coherent states in graphene.
\newblock {\em J. Phys.: Conf. Ser.}, 1194:012025, 2019.

\bibitem{Celeita2020}
M.~Castillo-Celeita and D.J. Fern\'andez.
\newblock Dirac electron in graphene with magnetic fields arising from
  first-order intertwining operators.
\newblock {\em J. Phys. A: Math. Theor.}, 53:035302, 2020.

\bibitem{fgo20}
David~J Fern{\'{a}}ndez~C, Juan~D Garc{\'{\i}}a~M, and Daniel O-Campa.
\newblock Electron in bilayer graphene with magnetic fields leading to shape
  invariant potentials.
\newblock {\em Journal of Physics A: Mathematical and Theoretical},
  53(43):435202, oct 2020.

\bibitem{fm20}
David~J Fern{\'{a}}ndez~C and Dennis~I. Mart\'inez-Moreno.
\newblock Bilayer graphene coherent states.
\newblock {\em The European Physical Journal Plus}, 135:739, sept 2020.

\bibitem{Andrianov1993}
A.A. Andrianov, M.V. Ioffe, and V.P. Spiridonov.
\newblock Higher-derivative supersymmetry and the witten index.
\newblock {\em Phys. Lett. A}, 174:273, 1993.

\bibitem{Andrianov1995}
A.A. Andrianov, M.V. Ioffe, F.~Cannata, and J.P. Dedonder.
\newblock Second order derivative supersymmetry, q deformations and the
  scattering problem.
\newblock {\em Int. J. Mod. Phys. A}, 10:2683, 1995.

\bibitem{Samsonov1999}
B.F. Samsonov.
\newblock New possibilities for supersymmetry breakdown in quantum mechanics
  and second-order irreducible darboux transformations.
\newblock {\em Phys. Lett. A}, 263:274, 1999.

\bibitem{Salinas2003}
David~J Fern{\'a}ndez~C and Encarnaci{\'o}n Salinas-Hern{\'a}ndez.
\newblock The confluent algorithm in second-order supersymmetric quantum
  mechanics.
\newblock {\em Journal of Physics A: Mathematical and General},
  36(10):2537--2543, feb 2003.

\bibitem{Salinas2005}
David~J. Fern{\'a}ndez~C. and Encarnaci{\'o}n Salinas-Hern{\'a}ndez.
\newblock Wronskian formula for confluent second-order supersymmetric quantum
  mechanics.
\newblock {\em Physics Letters A}, 338(1):13 -- 18, 2005.

\bibitem{Nicolas2005}
David~J. Fernandez~C. and Nicolas Fernandez-Garcia.
\newblock Higher-order supersymmetric quantum mechanics.
\newblock {\em AIP Conference Proceedings}, 744, 02 2005.

\bibitem{Fernandez2010}
D.J. Fernandez.
\newblock Supersymmetric quantum mechanics.
\newblock {\em AIP Conf. Proc.}, 1287:3, 2010.

\bibitem{Fernandez2019}
D.J. Fernandez.
\newblock Trends in supersymmetric quantum mechanics.
\newblock {\em Integrability, Supersymmetry and Coherent States, CRM Series in
  Mathematical Physics, Springer, Cham}, page~37, 2019.

\bibitem{Barnana2020}
David~J Fern{\'{a}}ndez~C and Barnana Roy.
\newblock Confluent second-order supersymmetric quantum mechanics and spectral
  design.
\newblock {\em Physica Scripta}, 95(5):055210, feb 2020.

\bibitem{Mermin1976}
Neil~W. Ashcroft and N.~David Mermin.
\newblock {\em Solid State Physics}.
\newblock Harcourt College Publishers, USA, college edition edition, 1976.

\bibitem{Saito1998}
R.Saito, G.~Dresselhaus, and M.S. Dresselhaus.
\newblock {\em Physical Properties of Carbon Nanotubes}.
\newblock Imperial College Press, London, first edition edition, 1998.

\bibitem{Raza2012}
Hassan Raza.
\newblock {\em Graphene Nanoelectronics: Metrology, Synthesis, Properties and
  Applications}.
\newblock Springer-Verlag, Berlin, first edition edition, 2012.

\bibitem{Wu2012}
Si~Wu, Matthew Killi, and Arun Paramekanti.
\newblock Graphene under spatially varying external potentials: Landau levels,
  magnetotransport, and topological modes.
\newblock {\em Physical Review B}, 85, 02 2012.

\bibitem{Ferreira2011}
Aires Ferreira, J.~Viana-Gomes, Johan Nilsson, E.~R. Mucciolo, N.~M.~R. Peres,
  and A.~H. Castro~Neto.
\newblock Unified description of the dc conductivity of monolayer and bilayer
  graphene at finite densities based on resonant scatterers.
\newblock {\em Phys. Rev. B}, 83:165402, Apr 2011.

\bibitem{Compean2005}
C~B Compean and M~Kirchbach.
\newblock The trigonometric rosen{\textendash}morse potential in the
  supersymmetric quantum mechanics and its exact solutions.
\newblock {\em Journal of Physics A: Mathematical and General}, 39(3):547--557,
  dec 2005.

\bibitem{Gangopadhyaya2018}
A.~Gangopadhyaya, J.~Mallow, and C.~Rasinariu.
\newblock {\em Supersymmetric Quantum Mechanics}.
\newblock World Scientific, Singapore, second edition, 2018.

\bibitem{Junker2019}
G.~Junker.
\newblock {\em Supersymmetric Methods in Quantum, Statistical and Solid State
  Physics}.
\newblock IOP Publishing Ltd, Bristol, second edition, 2019.

\bibitem{df11}
Samuel Dom\'inguez-Hern\'andez and David~J Fern{\'{a}}ndez~C.
\newblock Rosen–morse potential and its supersymmetric partners.
\newblock {\em International Journal of Theoretical Physic}, 50(7):1993--2001,
  2011.

\bibitem{mnr00}
B~Mielnik, LM~Nieto, and O~Rosas-Ortiz.
\newblock The finite difference algorithm for higher order supersymmetry.
\newblock {\em Physics Letters A}, 269(2-3):70--78, 2000.

\bibitem{fs11}
David~J Fern{\'a}ndez~C and Encarnaci{\'o}n Salinas-Hern{\'a}ndez.
\newblock Hyperconfluent third-order supersymmetric quantum mechanics.
\newblock {\em Journal of Physics A: Mathematical and General}, 44(36):365302,
  aug 2011.

\bibitem{bff12}
David Bermudez, David~J Fern{\'a}ndez~C, and Nicol\'as Fern\'andez-Garc\'ia.
\newblock Wronskian differential formula for confluent supersymmetric quantum
  mechanics.
\newblock {\em Physics Letters A}, 376(5):692--696, jan 2012.

\bibitem{cjp15}
Francisco Correa, Vít Jakubsky, and Mikhail~S. Plyushchay.
\newblock Pt-symmetric invisible defects and confluent darboux-crum
  transformations.
\newblock {\em Physical Review A}, 92:023839, aug 2015.

\bibitem{gq15}
Yves Grandati and Christiane Quesne.
\newblock Confluent chains of dbt: Enlarged shape invariance and new orthogonal
  polynomials.
\newblock {\em SIGMA}, 11:061, jul 2015.

\bibitem{cs15a}
Alonso Contreras-Astorga and Axel Schulze-Halberg.
\newblock The generalized zero-mode supersymmetry scheme and the confluent
  algorithm.
\newblock {\em Annals of Physics}, 354:353--364, mar 2015.

\bibitem{cs15b}
Alonso Contreras-Astorga and Axel Schulze-Halberg.
\newblock On integral and differential representations of jordan chains and the
  confluent supersymmetry algorithm.
\newblock {\em Journal of Physics A: Mathematical and Theoretical},
  48(31):315202, jul 2015.

\bibitem{be16}
David Bermudez.
\newblock Wronskian differential formula for k-confluent susy qm.
\newblock {\em Annals of Physics}, 364:35--52, jan 2016.

\bibitem{cs17}
Alonso Contreras-Astorga and Axel Schulze-Halberg.
\newblock Recursive representation of wronskians in confluent supersymmetric
  quantum mechanics.
\newblock {\em Journal of Physics A: Mathematical and Theoretical},
  50(10):105301, feb 2017.

\bibitem{sy18}
Axel Schulze-Halberg and Ozlem Yesiltas.
\newblock The generalized confluent supersymmetry algorithm: Representations
  and integral formulas.
\newblock {\em Journal of Mathematical Physics}, 59(4):043508, apr 2018.

\bibitem{Sakurai1994}
Jun~John Sakurai.
\newblock {\em {Modern quantum mechanics; rev. ed.}}
\newblock Addison-Wesley, Reading, MA, 1994.

\end{thebibliography}
\bibliographystyle{unsrt}

\begin{appendices}
\section{Relationship between the wavenumber $k$ and the potential parameter $\kappa$} \label{apendice}
In order to find the relationship between $\kappa$ and the wavenumber $k$, let us remember that 
\begin{equation}
k=\frac{1}{2}\left(\eta(x)-\frac{2e}{c\hbar}\mathcal{A}(x)\right),
\label{A1}
\end{equation}
where the function $\eta$ depends on $\kappa$, since $V_0(x)$ involves $\kappa$ and some other parameters denoted globally by the symbol $\gamma$. If we express $\eta$ as
\begin{equation}
\eta(x)=f(x;\gamma)+C_1(\gamma),
\label{A2}
\end{equation}
with $f$ being a function of $x$ and $C_1$ being a constant, we will get
\begin{equation}
k=\frac{1}{2}\left(f(x;\gamma)+C_1(\gamma)-\frac{2e}{c\hbar}\mathcal{A}(x)\right).
\label{A3}
\end{equation}
On the other hand, the magnetic field $\mathcal{B}$ and $\eta'$ are related by
\begin{align}
\mathcal{B}(x)&=\frac{c\hbar}{2e}\eta'(x) = \frac{c\hbar}{2e}f'(x;\gamma).
\label{A4}
\end{align}
But the vector potential $\mathcal{A}(x)$ is also related to $\mathcal{B}(x)$ in the way
\begin{align}
\mathcal{A}(x)&=\int \mathcal{B}(x)dx =\frac{c\hbar}{2e}\int f'(x;\gamma)dx =\frac{c\hbar}{2e}
\left(f(x;\gamma)+C_2\right),
\label{A5}
\end{align}
where $C_2$ is an integration constant in general different from $C_1$. By plugging equations~(\ref{A5}), (\ref{A2}) into equation~(\ref{A1}) we arrive at
\begin{equation}
k=\frac{1}{2}\left(C_1(\gamma)-C_2\right).
\label{A6}
\end{equation}
Thus, we just need to identify the constant $C_1(\gamma)$ of $\eta$ in order to get $k$, a task which in general is not easy because the integral in equation~\eqref{A5} cannot be always calculated in  a simple way. Furthermore, it is worth noticing that in the confluent case it could appear also a dependence on the parameter $\mathrm{w}_0$.

\section{Continuity equation} \label{apendice2} 

Concerning the assumptions made at section \ref{section2} to determine the second-order differential intertwining operator $L_{2}^{-}$, the Hamitonian in equation~(\ref{extra2}) can also be written as
\begin{equation} \label{B1}
H = \frac{1}{2m^{*}}
\left(\begin{array}{cc}
0 & \left(\pi - i\frac{e}{c}\mathcal{A}\right)^{2} - \hbar^{2}f \\
 \left(\pi + i\frac{e}{c}\mathcal{A}\right)^{2} - \hbar^{2}f & 0
\end{array}\right),
\end{equation}
where the Landau gauge has been chosen, but now we consider that $\mathcal{A}(x,y), f(x,y)$ are two real functions depending on both coordinates $x,y$ and $\pi = p_{x} - ip_{y}$. Since $p_{l} = -i\hbar\partial_{l}$, $l = x,y$ and supposing normalized vectors $\Psi = \left(\psi_{2}, \psi_{0}\right)^{T}$, with $\psi_{2}$ and $\psi_{0}$ being arbitrary complex functions of $x,y$, we get that
\begin{eqnarray} \label{B2}
\nonumber H\Psi & = \frac{\hbar^{2}}{2m^{*}}
\Big(\partial_{x}\left(-\partial_{x} + i\partial_{y}\right)\psi_{0} + \partial_{y}\left(\partial_{y} + i\partial_{x}\right)\psi_{0} - 2\frac{e\mathcal{A}}{c\hbar}\partial_{x}\psi_{0} + i2\frac{e\mathcal{A}}{c\hbar}\partial_{y}\psi_{0} \\
\nonumber & - \psi_{0}\partial_{x}\frac{e\mathcal{A}}{c\hbar} + i\psi_{0}\partial_{y}\frac{e\mathcal{A}}{c\hbar} - \frac{e^{2}\mathcal{A}^{2}}{c^{2}\hbar^{2}}\psi_{0} - f\psi_{0},\
\partial_{x}\left(-\partial_{x} - i\partial_{y}\right)\psi_{2} + \partial_{y}\left(\partial_{y} - i\partial_{x}\right)\psi_{2} \\
& + 2\frac{e\mathcal{A}}{c\hbar}\partial_{x}\psi_{2} + i2\frac{e\mathcal{A}}{c\hbar}\partial_{y}\psi_{2} + \psi_{2}\partial_{x}\frac{e\mathcal{A}}{c\hbar} + i\psi_{2}\partial_{y}\frac{e\mathcal{A}}{c\hbar} - \frac{e^{2}\mathcal{A}^{2}}{c^{2}\hbar^{2}}\psi_{2} - f\psi_{2}\Big)^{T}.
\end{eqnarray}
Now, by remembering that for the dynamical law 
\begin{equation} \label{B3}
i\hbar\frac{\partial\Psi}{\partial t} = H \Psi, 
\end{equation}
the continuity equation is given by 
\begin{equation} \label{B4}
\frac{\partial\rho}{\partial t} + \frac{i}{\hbar}\left[\Psi^{\dagger}(H\Psi) - (H\Psi)^{\dagger}\Psi\right] = 0,
\end{equation}
then the second term in the previous equation turns out to be 
\begin{eqnarray} \label{B5}
\nabla\cdot\frac{\hbar}{m^{*}}\Bigg\{\left[\mathrm{Im}\left(\Psi^{\dagger}j_{x}\Psi\right) + \frac{e\mathcal{A}}{c\hbar}\Psi^{\dagger}\sigma_{y}\Psi\right]\hat{e}_{x} + \left[\mathrm{Im}\left(\Psi^{\dagger}j_{y}\Psi\right) - \frac{e\mathcal{A}}{c\hbar}\Psi^{\dagger}\sigma_{x}\Psi\right]\hat{e}_{y}\Bigg\}.
\end{eqnarray}
Then, the current density in our case can be written as 
\begin{equation}  \label{B6}
\vec{J} = \frac{\hbar}{m^{*}}\left[\mathrm{Im}\left(\Psi^{\dagger}\vec{j}\Psi\right) + \frac{e\mathcal{A}}{c\hbar}\Psi^{\dagger}\vec{\varsigma}\ \Psi\right],
\end{equation}
where $j_{x} = \sigma_{x}\partial_{x} + \sigma_{y}\partial_{y}$, $j_{y} = \sigma_{y}\partial_{x} - \sigma_{x}\partial_{y}$, $\varsigma_{l} = \varepsilon_{lm}\sigma_{m}$, $l,m = x,y$, $\varepsilon_{lm}$ is the 2-dim Levi-Civita symbol and $\sigma_{m}$ are the Pauli matrices.

It is worth to notice that, since the Hamiltonian (\ref{extra2}) is quadratic in the momentum, the current density obtained is similar to the one presented in \cite{Sakurai1994}. Moreover, if an arbitrary gauge is taken (keeping the magnetic field along $z$-direction), the linear operators in the momentum which are off the diagonal of $H$ change to $\pi + (e/c)\mathcal{A}_{x} - i(e/c)\mathcal{A}_{y}$ and $\pi^{\dagger} + (e/c)\mathcal{A}_{x} + i(e/c)\mathcal{A}_{y}$. Therefore, the current density will acquire an extra term given by $(e\mathcal{A}_{x}/c\hbar)\psi^{\dagger}\vec{\sigma}\psi$, with $\vec{\sigma}$ being the vector whose elements are the Pauli matrices.
\end{appendices}

\end{document}